\documentclass[journal]{IEEEtran}

\usepackage{amsmath,amssymb}
\usepackage{mathptmx}
\usepackage{flushend}

\usepackage{graphicx}
\usepackage{epsfig}
\usepackage{epstopdf}
\usepackage[percent]{overpic}

\usepackage[normalsize, center]{subfigure}
\usepackage[font=small]{caption}
\usepackage{cite}

\usepackage{array}
\usepackage{multirow}
\usepackage{hhline}% http://ctan.org/pkg/hhline
\usepackage{diagbox}
\usepackage{booktabs}
\usepackage{makecell}
\usepackage[table]{xcolor}
\usepackage{color}

\def\docAuthor{M. Sezer Erk{\i}l{\i}nc{c}}
\def\docTitle{Comparison of Low Complexity Coherent Receivers for UDWDM-PONs ($\lambda$-to-the-user)}

\usepackage[pdftex,colorlinks=true,bookmarks=false, linkcolor=blue,citecolor=blue,urlcolor=blue,baseurl='',pdfstartview=FitH,pdftitle=\docTitle, pdfauthor=\docAuthor, pdfdisplaydoctitle=true]{hyperref}

\usepackage{color}

\newcommand{\etal}{\emph{et~al.\/}}
\newcommand{\eg}  {\emph{e.g.,\/}}
\newcommand{\ie}  {\emph{i.e.,\/}}

\begin{document}
\title{\docTitle}
% note positions of commas and non-breaking spaces ( ~ ) LaTeX will not break a structure at a ~ so this keeps an author's name from being broken across two lines.

\author{M.~Sezer~Erk{\i}l{\i}n\c{c},~\IEEEmembership{Member,~IEEE,}
		Domani\c{c}~Lavery,~\IEEEmembership{Member,~IEEE,}
        Kai~Shi,~\IEEEmembership{Member,~IEEE,}
        Benn~C.~Thomsen,~\IEEEmembership{Member,~IEEE,}
        Robert~I.~Killey,~\IEEEmembership{Senior Member,~IEEE,}
        Seb~J.~Savory,~\IEEEmembership{Fellow,~IEEE,~Fellow,~OSA,}
        and Polina~Bayvel,~\IEEEmembership{Fellow,~IEEE,~Fellow,~OSA}
\thanks{Manuscript  received  December  3,  2017;  revised  February 7, 2018. This work was supported by the EPSRC EP/J008842/1 and UNLOC EP/J017582/1.}
\thanks{M. Sezer Erk{\i}l{\i}n\c{c}, Domani\c{c} Lavery, Robert I. Killey, Polina Bayvel are with the Optical Networks Group, in the Dept. of Electronic and Electrical Engineering at University College London, London WC1E 7JE, UK. (e-mail: m.erkilinc@ee.ucl.ac.uk).}
\thanks{Kai Shi and Benn C. Thomsen were with the Optical Networks Group, and now are with the Microsoft Research Ltd, Cambridge, CB1 2FB, UK. (e-mails: t-kashi@microsoft.com; benn.thomsen@microsoft.com)}
\thanks{Seb J.~Savory is with the Dept. of Eng., Electrical Eng. Division, University of Cambridge, 9 JJ Thomson Avenue, Cambridge, CB3 0FA, UK. (e-mail: sjs1001@cam.ac.uk).}
\thanks{Colour versions of one or more of these figures in this paper are available online at http://www.ieeexplore.ieee.org.}}

\markboth{Journal of Lightwave Technology,~Vol.~XX, No.~X, Month~2018}
{Erk{\i}l{\i}nc{c} \MakeLowercase{\textit{et al.}}: Title}

\maketitle% make the title area

\begin{abstract}
It is predicted that demand in future optical access networks will reach multi-gigabit/s per user. However, the limited performance of the direct detection receiver technology currently used in the optical network units at the customers' premises restricts data rates per user. Therefore, the concept of coherent-enabled access networks has attracted attention in recent years, as this technology offers high receiver sensitivity, inherent frequency selectivity, and linear field detection enabling the full compensation of linear channel impairments. However, the complexity of conventional (dual-polarisation digital) coherent receivers has so far prevented their introduction into access networks. Thus, to exploit the benefits of coherent technology in the ONUs, low complexity coherent receivers, suitable for implementation in ONUs, are needed. In this paper, the recently proposed low complexity coherent (i.e., polarisation-independent Alamouti-coding heterodyne) receiver is, for the first time, compared in terms of its minimum receiver sensitivity with five previously reported receiver designs, including a detailed discussion on their advantages and limitations. It is shown that, {\color{red}of all the receiver configuration considered, the Alamouti-coding based receiver approach allows the lowest number of photons per bit (PPB) transmitted (with a lower bound of 15.5 PPB in an ideal implementation of the system) whilst requiring the lowest optical receiver hardware complexity (in terms of the number of components).} It also exhibits comparable complexity to the currently deployed direct-detection receivers, which typically require over 1000 PPB. Finally, a comparison of experimentally achieved receiver sensitivities and transmission distances using these receivers is presented. {\color{red}The highest spectral efficiency and longest transmission distance at the highest bit rate (10~Gb/s) reported using the Alamouti-coding receiver, which is also the only one, to date, to have been demonstrated in a full system bidirectional transmission.}
\end{abstract}
 

\begin{figure}[t]
	\centering
	\includegraphics[width=1\linewidth]{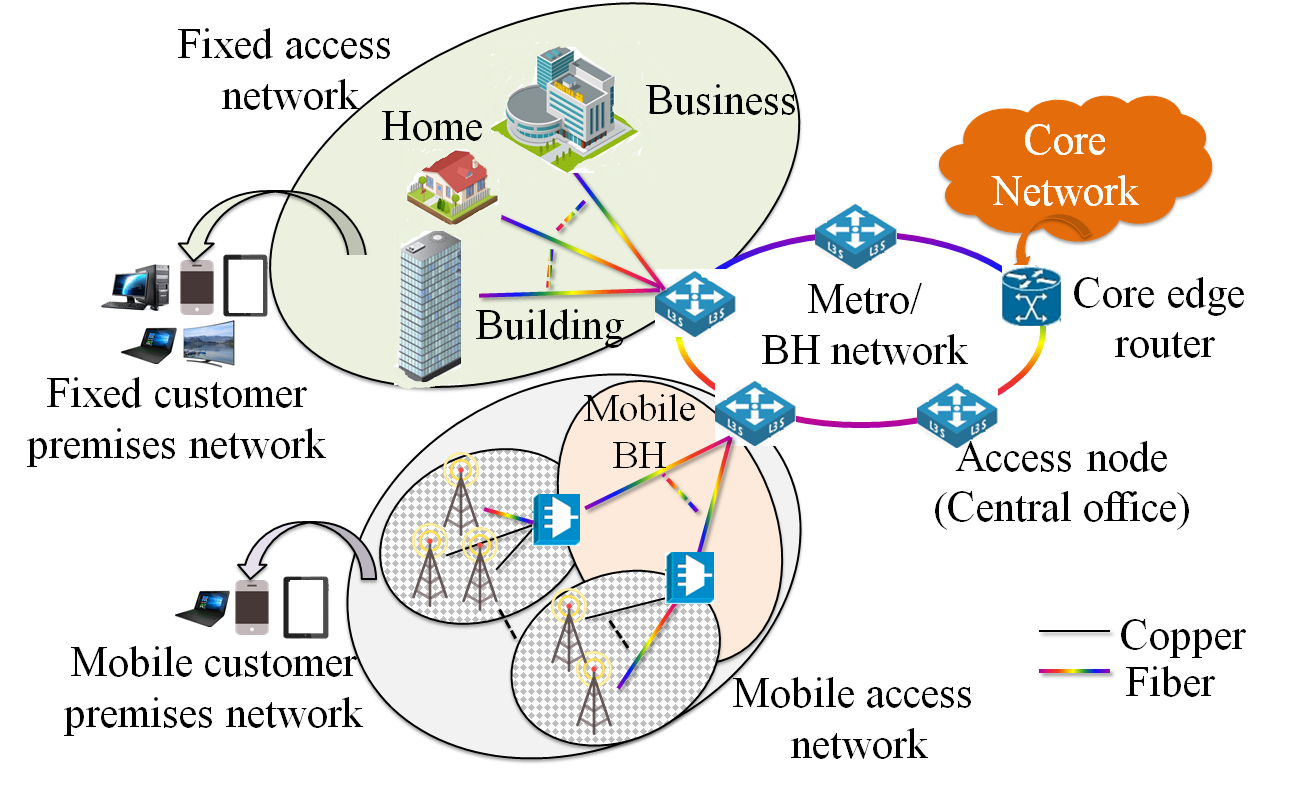}
	\caption{Simplified reference topology for optical access and mobile backhaul (BH) networks using a PON architecture with tree topology.}
	\label{Fig:Overview}   
\end{figure}

\section{Introduction}
The significant increase in the number and download speeds of mobile devices combined with new emerging mobile technologies, such as 5G, and data-intensive applications (\eg~high-definition video-on-demand, online entertainment, cloud computing/storage services, the Internet of Things, and Big Data) has resulted in continuously increasing demand for bandwidth and growing levels of data traffic at residential areas and businesses~\cite{Cisco17}. This demand will eventually reach multi-gigabit connection speeds per subscriber delivered via optical access networks; specifically, fibre-to-the-business/building/home/premises (FTTx) enabling high capacity and low latency connections. Additionally, the generated data in access networks needs to be backhauled to core networks\footnote{European Commission proposes to have access internet connections of download/upload speeds of $\geq$1~Gb/s for all schools, main public services and enterprises by 2025. Therefore, ubiquitous multi-Gb/s connectivity will be required for residential markets, and bandwidth demand for business and backhaul markets will be even exceeding by a factor of ten that of the residential markets~\cite{EC}.}, as illustrated in Fig.~\ref{Fig:Overview}. 

\begin{figure}[t]
	\centering
	\includegraphics[width=0.85\linewidth]{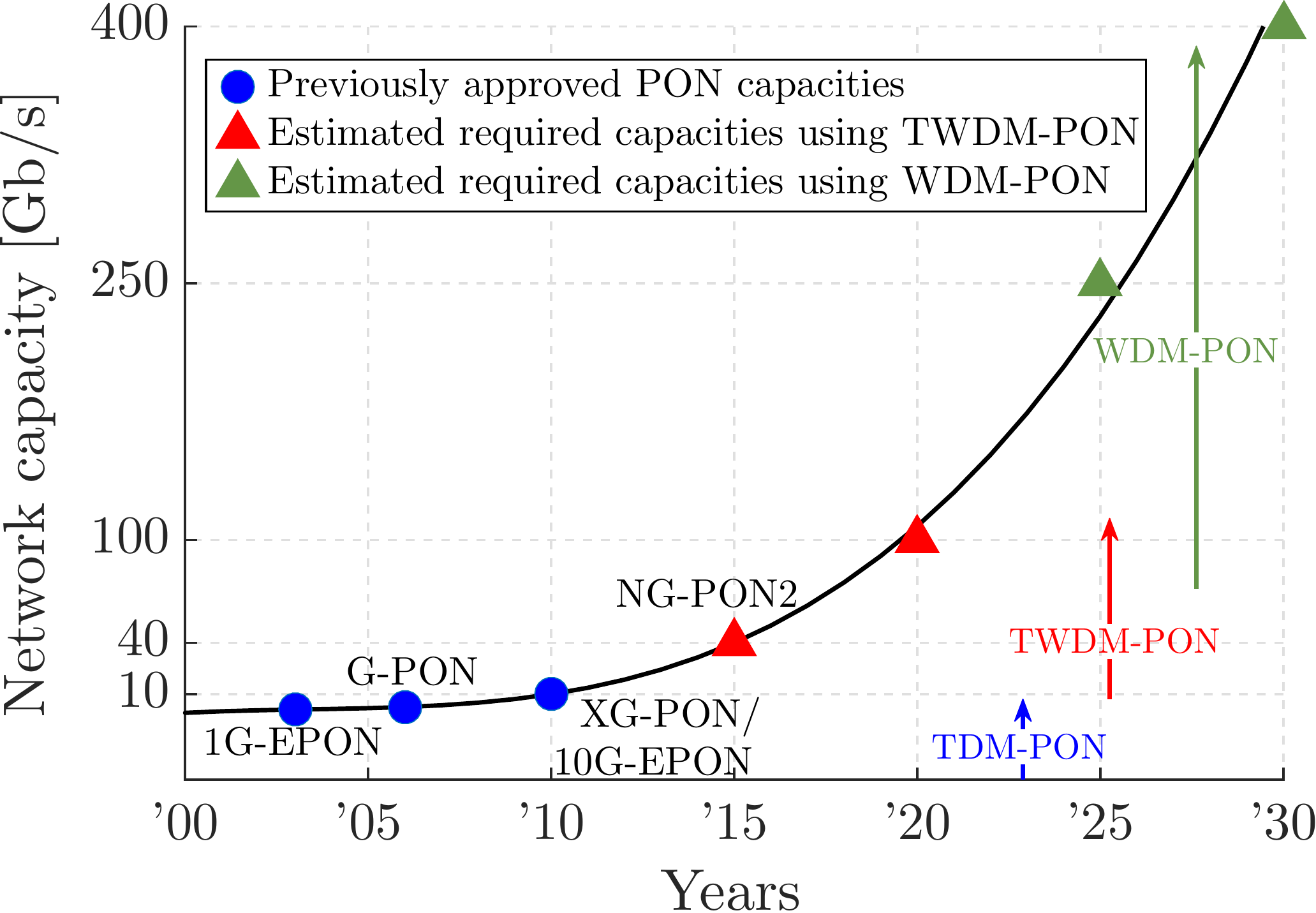}
	\caption{The capacity growth in optical access networks over time. The blue markers represent the previously approved PON standards whereas red and green markers are the projections based on the consented previous standards~\cite{Cisco17,ITU_NG-PON2, ITU_XGPON,IEEE_EPON}. The network capacity is estimated by the downstream speeds/user $\times$ number of users.}
	\label{Fig:DataRates}
\end{figure}

Optical fibre access systems based on passive optical networks (PONs) with a tree topology,~\ie~using only passive components in the physical infrastructure, are widely considered to be the best option to minimise the cost~\cite{Nesset_JOCN17, Huawei13}. Although currently employed signalling scheme, time-division multiple access (TDMA), in PONs offers a cost-effective solution for such networks, it comes at the expense of requiring transceivers with electrical bandwidths orders of magnitude higher than the bandwidth accessible by each individual subscriber. Therefore, the bandwidth limitations of transceiver electronics with TDMA signalling will make it challenging to provide multi-gigabit/s per subscriber. To overcome this limitation, ITU-T has recently standardized the second generation PON, referred to as NG-PON2~\cite{ITU_NG-PON2}, which exploits both the time and wavelength domains, offering an aggregate network capacity of 40~Gb/s. However, based on the current trends in capacity growth in access networks, it is forecast that the required (aggregate) capacity will exceed 100~Gb/s by 2020 reaching 250~Gb/s by 2025, as shown in Fig.~\ref{Fig:DataRates}. Thus, due to their high level of data rate scalability, wavelength division multiplexed (WDM) PONs are being considered by network operators and service providers~\cite{Kani10, Pachnicke16}.

% LookPachnicke16

In PONs, an asymmetric transceiver architecture is used; for the down-link, the limits on complexity for the transmitter in the optical line terminal (OLT) at the Central Office are less stringent than those for the receiver in the ONU, since the cost of the transmitter, which sends data to multiple ONUs, is shared by all the users supported in the network. In contrast, the cost of each ONU is born solely by the user, and hence, low complexity and low cost are more critical for the ONU. Direct detection (DD) receivers have, to date, been preferred by the operators over dual-polarisation digital coherent receivers due to their simpler architecture (with fewer optical components) offering higher laser linewidth tolerance without needing  complex digital signal processing (DSP). 

WDM-PONs use an arrayed waveguide grating (AWG) filter at the remote node to distribute the wavelengths to the end users, `colouring' the network~\cite{Bhar14}. Thus, the DD receivers used for TDM-PON can also be used for WDM-PONs which is particularly desirable. However, colouring the network using an AWG in the remote node reduces the network flexibility such as requiring a fixed wavelength for each user's down-link. {\color{red}Similarly, in the up-link, the requirement for wavelength diversity would lead to an inventory cost problem for ONUs using fixed wavelength lasers.} 

\renewcommand\thefootnote{\textcolor{red}{\arabic{footnote}}}

To enable flexible network operation by allowing filterless/colourless operation of the ONU without a midspan AWG or tunable optical filters in the ONU, coherent receivers (inherently wavelength/frequency selective) can be deployed, selecting a wavelength channel simply by tuning the local oscillator (LO) laser to the wavelength of the downstream channel of interest. Note that colourless operation is desirable for the ONU to make it cost-effective in volume production due to its standardized design, and more manageable for network operators by expediting its operation and maintenance\footnote{If all ONUs are identical (`colourless'), then fewer ONUs are required to be kept in inventories, ultimately reducing costs. {\color{red}However, achieving this by using tunable laser introduces an extra cost consideration~\cite{Grobe_JLT14}}.}. Fine wavelength selectivity in coherent-enabled WDM-PONs enables the use of (ultra-)dense wavelength spacing whilst requiring no sophisticated optical wavelength filtering. Recent demonstrations of this technique include 10~Gbps/$\lambda$ transmission using a 5~GHz grid~\cite{Lavery_OpEx10} and 3.75~Gbps/$\lambda$ using a 2.5~GHz grid~\cite{Ferreira_PTL17}.

\renewcommand\thefootnote{\textcolor{black}{\arabic{footnote}}}

In addition to this key advantage, coherent receivers offer significantly higher receiver sensitivities~\cite{Rohde_JLT14, Shahpari17} in comparison to DD receivers. This will be a major advantage in future PON technologies, operating at multi-Gb/s per subscriber, and offering higher loss budgets, enabling higher split ratios (\ie~increased number of users) and longer reach. The high receiver sensitivity enables high power budgets that can be shared arbitrarily between reach and split ratio depending on the network requirements\footnote{It is important to note that long-reach PONs can further reduce the operational cost by consolidating the backhaul and access network fibres, requiring no reach extenders or optical amplifiers at remote nodes~\cite{Shea_JLT07}.}. High split ratios reduce per-user costs since more end users can be supported in an access network using a single feeder fibre, and the principal multiplexing technique is expected to be WDM access (WDMA), rather than TDMA, avoiding the limitation on data rate per user that is imposed by the need for high bandwidth electronics with TDMA. Besides their higher sensitivity, the linear detection of coherent receivers, and the resulting ease of digital dispersion compensation, offer further cost benefits. Thus, the use of coherent reception and WDMA will be the key enabling factors to meet future optical access network demands.

Although digital coherent technology offers significant advantages, the complexity and high cost of conventional (polarisation- and phase-diverse) intradyne digital coherent receivers has prevented their use in PON applications. It is likely that future PONs will have similar cost constraints to those of today's PONs. Thus, low complexity (simplified) coherent technology can play a major key role in future access and mobile backhaul PONs. They can offer power budgets of around 40~dB which is approximately 10~dB higher than the NG-PON2 budget requirements. {\color{red} Assuming a 3.5~dB loss per 1:2 split and 0.25~dB/km fibre attenuation, this gain potentially leads to a two or four-fold increase in the number of users simultaneously with a corresponding transmission distance increase of 25 or 10 km, respectively.} 

\begin{table*} [t] 
	\small
	\centering
	\caption{Theoretical required photons per bit (PPB) (calculated from the required SNRs) at the HD-FEC threshold of BER=$4\times10^{-3}$, achievable using optically ideal pre-amplified coherent receivers. The error probability of the power-efficient modulation formats presented in this table can be found in~\cite[Ch.4, p.190]{Proakis08}. SE: Spectral efficiency.} 
	\label{Tab:RequiredPPBs}
	\renewcommand{\arraystretch}{1.1}% Tighter
	\newcolumntype{C}[1]{>{\centering\arraybackslash}m{#1}}
	\begin{center}
		\vspace{-1pc}
		\resizebox{\textwidth}{!}{
			\begin{tabular}{ C{2.8cm}  C{1.7cm} C{1.5cm} C{1.8cm}  C{1.3cm} C{1.5cm} C{1.4cm}  C{1.5cm}  C{1.2cm} }
				\toprule \toprule    
				
				\textbf{Format} & PSwitch-QPSK & DP-BPSK & DP-QPSK & DBPSK & 16-PPM & 4-PPM & OOK (2-PAM) & 4-PAM \\ 
				\midrule
				
				\textbf{Theoretical required PPB} & 2.9  & 3.5  &  3.5 &  4.1  & 4.1 & 6.9  & 7 & 23.9  \\ [0.5ex] 
				
				\textbf{Achievable SE (b/s/Hz)} & 3 & 2 &  4 & 1  & 0.25	& 0.5&  1 & 2\\ [0.5ex] 
				\bottomrule \bottomrule
		\end{tabular}}
	\end{center}
	\vspace{-1pc}
\end{table*}

Significant advances have recently been made in low complexity coherent receiver technology, for example, demonstrations employing vertical-cavity surface-emitting lasers combined with analogue signal processing. However, it is acknowledged that achieving polarisation-independent operation to minimise complexity, instead of polarisation-diversity as it comes at the expense of significant optical complexity, is an open research problem~\cite{Jensen_JLT14}. Hence, if polarisation-independent reception can be realised while avoiding the requirement for an optical polarisation tracking unit in the receiver, the complexity can be significantly reduced. To date, there are six reported low complexity polarisation-independent coherent receiver architectures employing various techniques. All these systems sacrifice one polarisation state or polarisation-diversity in the signalling but they do not require polarisation tracking between the signal and LO laser. A key question, then, is which low-complexity receiver designs offer the optimum solutions for future access and mobile backhaul PON applications in terms of capacity and reach, under a complexity constraint.

In this paper, initially, power-efficient modulation formats are discussed for applications in access networks, together with the proposed polarisation-independent high sensitivity yet low complexity coherent receiver designs. {\color{red}Following this, a numerical analysis of the shot noise limit of the recently proposed polarisation-independent coherent receiver, implemented using a polarisation-time block coding scheme combined with heterodyne reception~\cite{Erkilinc_ECOC15}, is presented, and its sensitivity performance is compared with those proposed by other research groups in terms of photons-per-bit~\cite{Glance87, Cano_ECOC15, Cano_ECOC14, Ciaramella_PTL14, Tabares_OFC17}. Additionally, the performance of two recently proposed simplified coherent receivers, exhibiting the lowest possible optical complexity in terms of required optical components, comparable to that of a direct detection receiver, is assessed in the presence of local oscillator relative intensity noise (LO-RIN). Finally, the experimentally achieved receiver sensitivities, power budgets, and transmission distances using these receivers are compared for the first time, and their relative hardware complexity requirements are discussed in detail.}
  
\begin{figure*}[t]
	\centering
 	\includegraphics[width=1\linewidth]{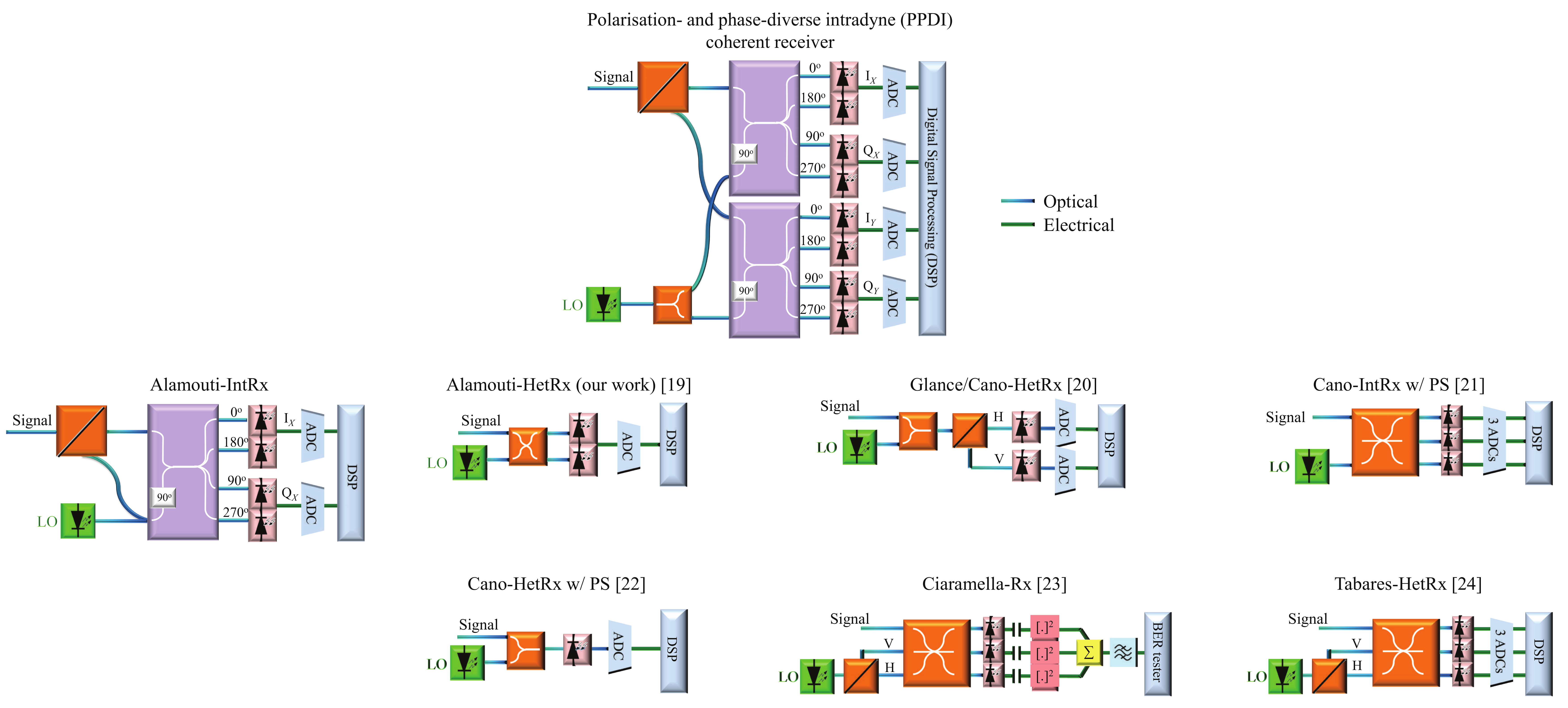}
 	\caption{The architecture of conventional polarisation- and phase-diverse intradyne and low complexity coherent receivers.}
 	\label{Fig:PICohRxDesigns}
 \end{figure*}

\section{High sensitivity, low complexity coherent detection in access networks}
\subsection{Power-efficient modulation formats for high sensitivity}
In access networks, receiver sensitivity at a targeted hard decision forward error correction (HD-FEC) limit is a key performance metric governing the system power budget, which determines the number of users that can be supported in a network and the transmission distance in optical, unamplified links, such as those in PONs. Therefore, there is an ongoing effort in optical communications research to realise power-efficient modulation formats to improve the receiver sensitivity.

%The process of photodetection inherently adds electrical noise to the as a result of (i) the random arrival time of the photons to the photodetector (\ie~the statistical nature of the photon to electron conversion), termed shot noise, and the random thermal motion of electrons in the load resistor in the front-end of an optical receiver, inducing fluctuations in the current generated by the photodiode, termed thermal noise.% %This is the material I teach to the second years, we kind of know that!!%
Considering the Poisson statistics, the quantum (upper) limit of photodetection for an on-off keying (OOK) signal, detected by an ideal direct detection receiver (neglecting thermal noise and dark current, and assuming 100\% quantum efficiency), is 2.4~photons per bit (PPB) at a bit error ratio (BER) of $4\times10^{-3}$~\cite[Ch.5, p.165]{Agrawal05}. However, this can only be achieved if the receiver uses an ideal optical pre-amplifier\footnote{Ideal pre-amplification for an optical receiver is modelled using an optical amplifier with a noise figure of 3~dB (setting the spontaneous emission factor $n_{sp}$ to 1) so that the signal-to-noise ratio (SNR) per bit becomes equal to the number of PPB~\cite[Ch.4, p.131]{Liu08}.} followed by an ideal (Nyquist) optical filter and cooled to a temperature near absolute zero, which is not practical for access networks. Thus, most practical direct detection receivers operate away from the quantum limit by $\geq$20~dB, with sensitivities exceeding 1000~PPB. On the other hand, shot noise imposes the quantum limit to receiver sensitivity for ideal optically pre-amplified coherent receivers. 

The theoretical shot noise limits for various coherently detected modulation formats, namely binary phase shift keying (BPSK), dual-polarisation quadrature PSK (DP-QPSK), differential BPSK (DBPSK), polarisation-switched (PSwitch) QPSK, 4- and 16-pulse position modulation (PPM), and 2- and 4-pulse amplitude modulation (PAM), are listed in Table~\ref{Tab:RequiredPPBs} in terms of PPB (power sensitivity in Watts normalised to the achievable bit rate) at the HD-FEC threshold of $4\times10^{-3}$. If high sensitivity is the absolute primary requirement (neglecting the spectral-efficiency), high-order $M$-PPM (\eg~M$\geq$16) is a clear choice, and, hence, is commonly used in optical free-space communications. However, it requires an $M$-fold increase in bandwidth compared to OOK (2-PAM), at a given bit rate. Although there are some proposed modulation formats enabled by stacking the formats, such as 16-PPM with DP-QPSK~\cite{Liu11} and 64-PPM with PSwitch-QPSK~\cite{Ludwig15} in which the lower bounds on sensitivity are 2.2 and 2.3~PPB, respectively, at the BER of $4\times10^{-3}$, they have yet to be demonstrated operating at data rates beyond a few Gb/s due to the poor bandwidth efficiency of the underlying PPM format. Thus, they are not favourable for future optical access networks. 

On the other hand, 4-PAM offers double the information per symbol compared to OOK, whilst requiring approximately three times the number of PPB due to the significant decrease in minimum Euclidean distance between symbols at a given signal power. When coherent detection using the conventional polarisation- and phase-diverse intradyne (PPDI) coherent receiver is considered, PSwitch-QPSK stands out as having the lowest required number of PPB for an uncoded transmission due to the largest possible Euclidean distance between symbols~\cite{Agrell09} at the HD-FEC threshold whereas there is a 0.6~PPB sensitivity difference between DP-QPSK and DBPSK due to the optimum constellation coding, as given in Table~\ref{Tab:RequiredPPBs}. PSwitch-QPSK requires 0.6 fewer PPB than DP-QPSK at the expense of offering a 25\% lower spectral efficiency. Thus, a trade-off between sensitivity and optical/electrical bandwidth requirements for the targeted capacity needs to be evaluated. The sensitivity difference between PSwitch- and DP-QPSK depends on the pre-FEC BER requirement,~\ie~it decreases when the pre-FEC BER increases, as discussed in detail in~\cite{Lavery_OpEx11}. It should be noted that it is desirable to use as low a FEC overhead as possible in PONs to reduce the power consumption and latency. Besides this, if the polarisation-diversity is sacrificed to implement a low complexity coherent receiver, the implementation of PSwitch-QPSK is not possible whereas single-polarisation QPSK might be a reasonable choice. 

The PPB values given in Table~\ref{Tab:RequiredPPBs} can be reached using optical pre-amplification, however, the use of optical pre-amplification is prohibitive in an ONU for reasons of cost and, potentially, safety. On the other hand, coherent detection offers significant sensitivity gains even without pre-amplification (achieving sensitivities close to the shot noise limit) since the achievable receiver sensitivity is determined by the LO power gain in coherent receivers. For instance, DD receivers require in the range of thousands of PPB to achieve a throughput of 10~Gb/s or higher, whereas coherent receivers require in the range of just tens of PPB. Thus, low complexity coherent receivers that are capable of detecting signals with power-efficient modulation formats, namely QPSK, DBPSK and OOK, are attractive for PON applications, as discussed in the next section.

\subsection{Description of the low complexity coherent receivers}\label{subsec:PICohRx_description}
\begin{figure}[t]
	\centering
	\includegraphics[width=1\linewidth]{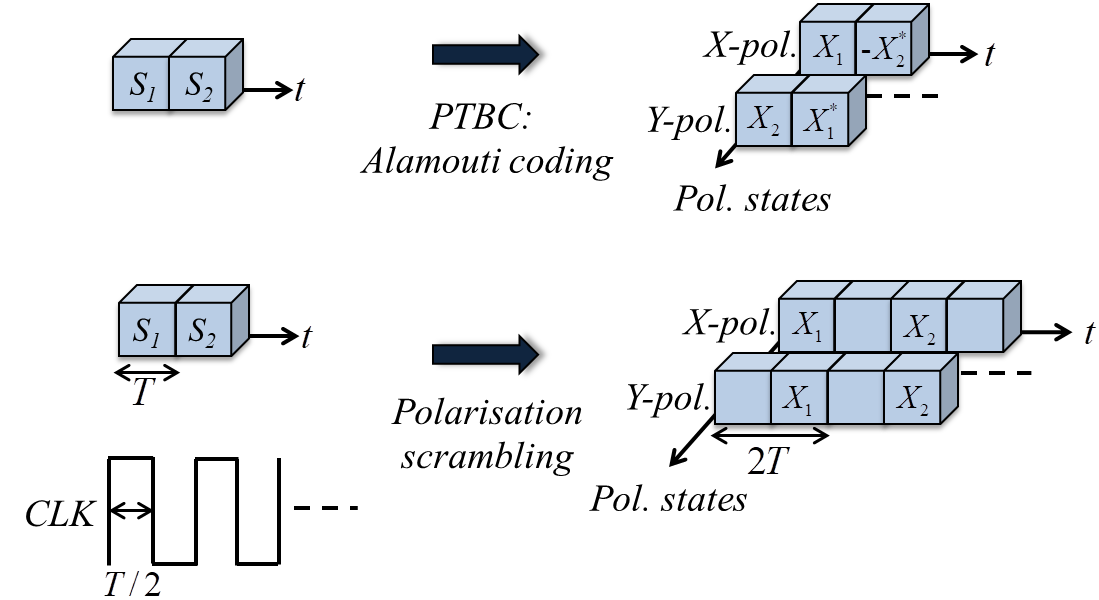}
	\caption{Schematic of the Alamouti-coding (top) and polarisation scrambling (bottom), yielding polarisation-independent operation.}
	\label{Fig:PolIndOp}
\end{figure}

The Alamouti receiver, first adopted for optical fibre communications by Shieh~\cite{Shieh08}, is a single polarisation coherent receiver which detects an Alamouti polarisation-time block coded (PTBC) signal, avoiding the requirement for an optical polarisation tracking unit in the receiver. In the event of polarisation rotation occurring along the fibre link, the coding scheme described in detail in~\cite{Erkilinc_JLT16, Faruk_OpEx16} enables the transmitted signal to be successfully recovered independently of the signal or LO state of polarisation. Although it introduces 50\% redundancy (half-rate coding, hence halving the achievable spectral efficiency) due to the replication of the transmitted symbols, as illustrated in Fig.~\ref{Fig:PolIndOp}, it leads to a significant simplification in the design compared to the conventional PPDI coherent receiver. The polarisation rotators/beam splitters (PBS), two of the balanced photodiodes (BPDs) and two of the analogue-to-digital converters (ADCs) of the PPDI receiver are no longer required, as depicted in Fig.~\ref{Fig:PICohRxDesigns} and referred to as Alamouti-IntRx. Moreover, by combining the coding scheme with heterodyne detection, an additional BPD followed by an ADC, and the 90$^{\circ}$ optical hybrid can be removed, as illustrated in Fig.~\ref{Fig:PICohRxDesigns} and referred to as Alamouti-HetRx. Despite the 3~dB penalty due to heterodyne detection, one of this technique's key advantages, in comparison to the other low complexity receivers using intradyne reception, is that it allows the simultaneous use of an ONU laser both as source and local phase reference in single fibre bidirectional links, as demonstrated in recent transmission experiments over installed fibre links in~\cite{Erkilinc_ECOC16, Erkilinc_NatureComm17}. In contrast, a photonic mixer (to shift the LO laser frequency) or a second laser is required to generate the upstream signal, which is highly undesirable, if homodyne/intradyne reception is employed.  

Glance~\etal~have proposed a polarisation-independent optical heterodyne receiver, consisting of a 3-dB coupler and a polarisation beam splitter (PBS), followed by two single-ended photodiodes (PDs), each detecting a polarisation component~\cite{Glance87}, as shown in Fig.~\ref{Fig:PICohRxDesigns}. Following signal detection, the photocurrents at intermediate frequencies are first filtered, demodulated separately, and finally, summed to obtain a baseband signal. Due to its capability of detecting two polarisation modes, the detection process is independent of the polarisation state of the received optical signal. Alternatively, phase diversity in time,~\ie~sending DBPSK symbols in alternating phases such as in-phase (I) and quadrature (Q) components in consecutive bits, has been proposed in~\cite{Cano_ECOC15-Glance}. This receiver structure is referred to as Glance/Cano-HetRx in the rest of the paper.

Moreover, Cano~\etal~\cite{Cano_ECOC14,Cano_ECOC15} proposed alternative low complexity coherent receiver designs for use in ONUs, achieving polarisation-independent detection using a polarisation scrambling (PS) method. In both studies, a polarisation synchronous intra-symbol scrambling technique, first introduced by Zhou and Caponio~\cite{Zhou94}, is utilised, in which every symbol is transmitted twice, in orthogonal polarisation states, during two time slots, as illustrated in Fig.~\ref{Fig:PolIndOp}. It is applied in the transmitter (optical line terminal) side using a polarisation modulator operating at twice the symbol (\textit{CLK}) rate by applying polarisation switching in the optical domain. Two low complexity coherent receivers using the PS technique (employing intradyne and heterodyne detection) have been demonstrated, and are referred to herein as Cano-IntRx~\cite{Cano_ECOC15} and Cano-HetRx~\cite{Cano_ECOC14}, respectively. The Cano-IntRx consists of a symmetric 3$\times$3 (1:1:1) coupler (using only two input ports) followed by three single-ended PDs and three ADCs, as illustrated in Fig.~\ref{Fig:PICohRxDesigns}~\cite{Xie_ECOC11}. In contrast, the Cano-HetRx has a simpler architecture which comprises a 3-dB coupler and a single-ended PD followed by a single ADC, as depicted in Fig.~\ref{Fig:PICohRxDesigns}. 

Finally, Ciaramella~\cite{Ciaramella_PTL14} has proposed a simplified coherent receiver achieving polarisation-independent reception. It employs a PBS and a symmetric 3$\times$3 coupler (utilising all three ports) followed by three single-ended PDs, as depicted in Fig.~\ref{Fig:PICohRxDesigns}. The LO laser is separated into two orthogonal states of polarisation (denoted as `H' and `V' in Fig.~\ref{Fig:PICohRxDesigns}) using a PBS, and subsequently, they are mixed with the signal component. The output photocurrents are passed through the DC-blocks, and then squared and summed to obtain the baseband signal. Finally, the signal is low-pass filtered before being input to a clock and data recovery circuit. The key advantage of the Ciaramella-Rx is that it requires only simple analogue processing,~\ie~there is no need for an ADC or DSP. However, this receiver design is limited to amplitude-shift keying (ASK) (\eg~OOK or 4-PAM) signalling and its tolerance to chromatic dispersion is lower than the other proposed receivers since the receiver linearity is lost due to squaring operation after the detection. A further disadvantage of this approach is that the receiver requires a large signal-LO frequency offset ($0.9\times$symbol rate) to avoid interference from low frequency components of the directly detected signal, and hence, the use of a single laser in the ONU, operating as both the upstream signal source and the downstream signal LO, is not possible. To regain the phase diversity and ability to use the ONU laser as a downstream LO and an upstream source laser enabled by heterodyne detection, Tabares~\etal~\cite{Tabares_OFC17} proposed modifying the Ciaramella-Rx by replacing the squaring operation with the linear combination of the three output photocurrents to remove the direct detection terms (identical to Cano-IntRx w/ PS), as explained in~\cite{Xie_ECOC11} whilst employing the same optical front-end design as the Ciaramella-Rx, as shown in Fig.~\ref{Fig:PICohRxDesigns}. Although the linear operations can be performed in the analogue domain, it should be noted that it is preferable for them to be carried out digitally, requiring 3 ADCs, to achieve high receiver sensitivity. The performance of these low complexity receivers and the sensitivity penalties due to such simplifications are discussed in detail in Section~\ref{sec:SimResults}.

\section{Numerical Simulations of Low Complexity coherent receivers}\label{subsec:NumSims}
\begin{figure}[t]
	\centering
	\includegraphics[width=0.9\linewidth]{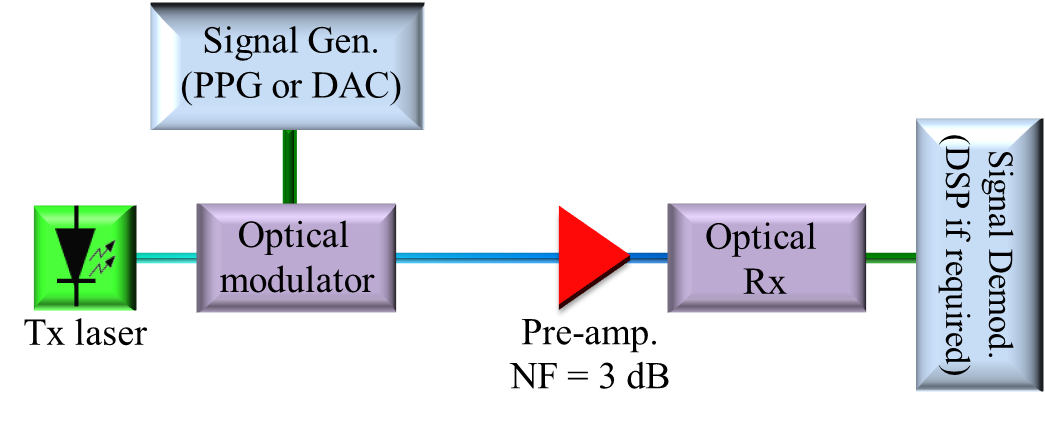}
	\caption{The block diagram representation of simulations. The modulators are DP-IQ and single-drive MZ modulators whereas the optical receivers are the low complexity coherent receivers, shown in Fig.~\ref{Fig:PICohRxDesigns}. NF: Noise figure.}
	\label{Fig:SimModel}
\end{figure}

Back-to-back numerical simulations were carried out to obtain estimates for the theoretical shot noise limits for the low complexity coherent receivers discussed in this paper (shown in Fig.~\ref{Fig:PICohRxDesigns}). The block diagram of the simulated system is shown in Fig.~\ref{Fig:SimModel}. It should be noted that ideal optical and electrical components were used in this simulation setup,~\eg~transmitter (source) lasers with negligible linewidth, optical modulators with a linear transfer function, optical couplers with ideal splitting ratios, an Erbium-doped fibre amplifier (EDFA) with a noise figure of 3~dB, and digital-to-analogue/analogue-to-digital converters (DACs/ADCs) without quantisation noise. In the transmitter, the electrical fields of the signals were modulated using mutually decorrelated de Bruijn bit sequences of length $2^{19}$ and, subsequently, they were oversampled by a factor of 8 to expand the simulation bandwidth for the realisation of heterodyne detection. A DP IQ-modulator was used to generate the optical DP-QPSK, Alamouti-coded single-carrier and OFDM QPSK and polarisation-scrambled DBPSK signals operating at 2.675, 5.35 and 10.7~GBaud (all corresponding to the bit rate of 10.7~Gb/s), respectively, whereas a single-drive Mach-Zehnder modulator was used to obtain 10.7~Gb/s optical DBPSK and OOK signals. The DSP used to generate the DP single carrier and OFDM QPSK signal from the bit sequences (with and without Alamouti coding) are explained in~\cite{Erkilinc_JLT16} and~\cite{Faruk_OpEx16}, respectively.

The receiver comprised an optical front end, ADC(s), DSP for signal demodulation (if required) and BER estimation, with each configuration shown in Fig.~\ref{Fig:PICohRxDesigns} being simulated. The relative intensity noise (RIN) of the LO was neglected, and the power and linewidth of the LO laser were assumed to be 20~dBm and 0~Hz, respectively. The thermal and shot noise for the photodiodes were modelled using the equations given in~\cite[Ch.4, p.151]{Agrawal10}, in which the absolute temperature and resistor load were assumed to be 300~K and 50~Ohm. In the case of Alamouti-IntRx and Alamouti-HetRx, the common-mode rejection ratio was assumed to be infinite for the balanced photodiode (BPD) and the quantum efficiency was assumed to be 1 for all photodiodes used in the receivers. Following the photodetection (and down-conversion which was required when heterodyne reception was considered), the DBPSK and OOK electrical signals were resampled to one sample-per-symbol, and subsequently, de-mapped to bits, while the DSP used for demodulation of the Alamouti-coded single carrier and OFDM QPSK signals is described in~\cite{Erkilinc_JLT16} and~\cite{Faruk_OpEx16}, respectively. Finally, the BER calculation was performed by hard-decision-based error counting over $2^{19}$ bits for all the modulation schemes.

\section{Simulation Results}\label{sec:SimResults}
This section initially presents the shot noise limits, obtained from numerical simulations, for the recently proposed low complexity coherent receiver, Alamouti-Rx. {\color{red} The impact on the shot noise limit as the simplifications are applied to the architecture of conventional polarisation- and phase-diverse intradyne (PPDI) coherent receiver is quantified. Following this, the performance of the Alamouti-Rxs (both Alamouti-IntRx and Alamouti-HetRx) is compared with that of the other proposed low complexity coherent receivers shown in Fig. 3. This is followed by a discussion on the impact of LO power for each of the receivers.

Finally, the performance of two polarisation-independent heterodyne receivers (employing Alamouti-coding and polarization synchronous intra-symbol scrambling), which exhibit minimum optical complexity (comprising a 3~dB coupler to combine signal and LO, followed by a single-ended PD, an ADC and DSP) for a coherent receiver are compared.}  

\subsection{Analysis of shot noise limit for the Alamouti-Rxs}
\begin{figure}[t]
	\centering
	\includegraphics[width=1\linewidth]{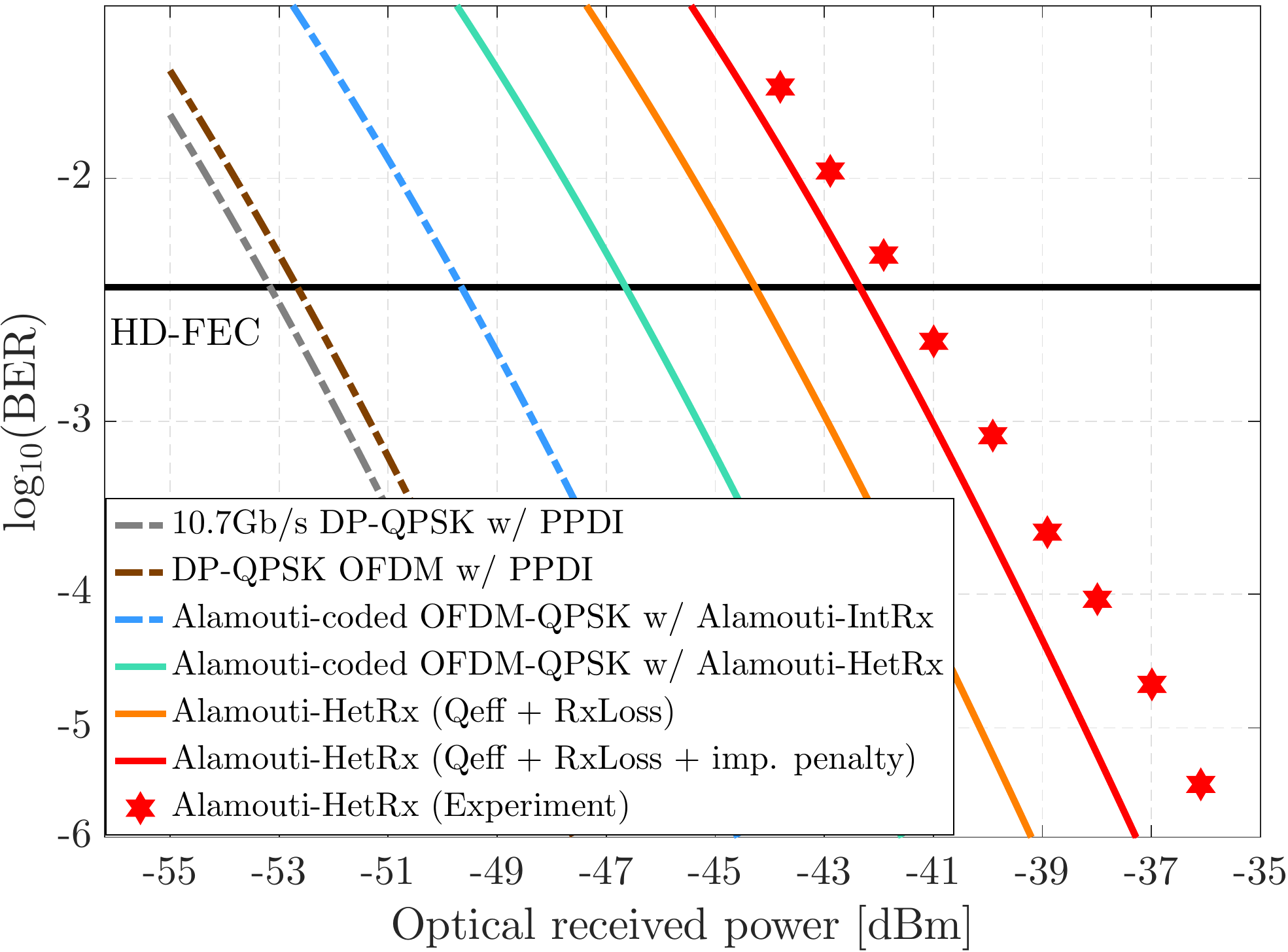}
	\caption{The effect in shot noise limit from the ideal conventional polarisation- and phase-diverse intradyne coherent receiver to the Alamouti-HetRx including the practical limitations. Experimental parameters are taken into account in practical simulations, and finally, they are validated experimentally.}
	\label{Fig:ShotNoise_PPDI2Alamouti}
\end{figure}

In this section, the simplification of the conventional PPDI coherent receiver to implement the Alamouti-Rx, and the resulting impact on the shot noise limits, is investigated in semi-numerical simulations. Initially, ideal system operation is considered,~\ie~neglecting quantisation noise and insertion losses, and assuming ideal splitting ratios. DP-QPSK signalling is used to realise the Alamouti polarisation-time block coding (PTBC) scheme. As a benchmark, the receiver sensitivity for a 10.7~Gb/s single carrier DP-QPSK signal detected using the PPDI coherent receiver was found to be -53.1~dBm at the HD-FEC threshold of $4\times10^{-3}$, as shown in Fig.~\ref{Fig:ShotNoise_PPDI2Alamouti}. A penalty of 0.3~dB was found for 10.7~Gb/s orthogonal frequency division multiplexed (OFDM) DP-QPSK due to the requirement for a small frequency guard band around DC frequencies. Alamouti PTBC comes at the price of a 3~dB penalty, inherent to the half-rate coding scheme,~\ie~sacrificing one polarisation to achieve polarisation-independent detection. Moreover, heterodyne detection results in a real-valued double sideband electrical signal (comprising signal and image bands) and doubles the energy of in band-noise compared to homodyne/intradyne detection,~\ie~the quantum-mechanical vacuum fluctuations in the image band are merged into the signal band appearing at the intermediate frequency. This causes an additional 3~dB penalty so that the theoretical shot noise limit for the 10.7~Gb/s Alamouti-coded OFDM-QPSK signal detected using the ideal Alamouti-HetRx becomes -46.8~dBm, as shown in Fig.~\ref{Fig:ShotNoise_PPDI2Alamouti}. 

Following this, the experimental parameters for the practical Alamouti-HetRx were considered in simulations. First, the quantum efficiency of the BPD was set to a value of 0.4 (whereas it was assumed to be 1 in the previous ideal system simulations, yielding a responsivity of 1.24~A/W). This results in a 0.9~dB sensitivity penalty. Note that the relationship between the quantum efficiency and responsivity is explained in~\cite{Kikuchi_JLT08}. Subsequently, the QPSK-OFDM Alamouti-HetRx was experimentally implemented using discrete optical components, introducing an extra 1.5~dB insertion loss to optimise the BPD common mode rejection ratio (CMRR), as shown in Fig.~\ref{Fig:ShotNoise_PPDI2Alamouti}. An optical delay line with an insertion loss of 1~dB was used to align the two input ports of the BPD in time, and attenuate the higher power BPD port by 0.5~dB to balance the power. Nonetheless, this loss can potentially be eliminated if the receiver is monolithically integrated. Furthermore, an additional combined implementation penalty of 2.4~dB was observed due to the use of a pilot tone added to the signal to correct symbol timing (phase) offset (0.5~dB), the cyclic prefix used to compensate for the chromatic dispersion from the standard single mode fibre span (SSMF) of up to 120~km (0.6~dB), the signal waveform clipping applied due to the limited DACs' resolution (0.7~dB), and the optical carrier with a carrier-to-signal power ratio of -8~dB inserted (by biasing the IQ modulator) in the transmitter for carrier phase recovery (0.6~dB). It should be noted that the optical carrier was inserted on both polarisation states of the Alamouti-coded OFDM signal to avoid power fading due to polarisation rotation. A receiver sensitivity of -41.6~dBm was experimentally achieved using the Alamouti-HetRx employing a 100~kHz linewidth  external cavity laser as a LO laser, as indicated by the markers in Fig.~\ref{Fig:ShotNoise_PPDI2Alamouti} and originally reported in~\cite{Erkilinc_JLT16}. Finally, the first demonstration of a full passive optical network over an installed fibre link (with the associated realistic fibre parameters such as loss, dispersion, polarisation mode dispersion, and polarisation rotation) was reported in~\cite{Erkilinc_NatureComm17} using the Alamouti-HetRx in a bidirectional operation (downlink speeds of 10.7~Gb/s and 21.4 Gb/s using higher-order QAM signal and an uplink speed of 10.7 Gb/s using BPSK signal). Crucially, the ONU laser in ~\cite{Erkilinc_NatureComm17} was used simultaneously as the LO laser for the downstream signal and the transmitter laser for the upstream signal, removing the requirement for an extra laser in the ONU, and making the ONU complexity comparable to current PON ONU technology. A discrepancy of 0.4~dB in optical received power between the simulations and experiment is due to the noise introduced by the transimpedance amplifier, but nonetheless, the experimentally measured receiver sensitivities show a good agreement with the practical system simulations.

\subsection{Comparison of shot noise limits for low complexity coherent receivers}
Next, the shot noise limit of the low complexity coherent receivers are compared. Ideal transceiver implementations, and a power-efficient modulation format are considered (single carrier QPSK in all cases except for the Ciaramella-Rx, in which OOK is used, since phase diversity is lost in this architecture as explained in Section~\ref{subsec:PICohRx_description}). The LO power was set to a constant 20~dB higher than the received signal power in all cases. 

\begin{figure}[t]
	\centering
	\includegraphics[width=1\linewidth]{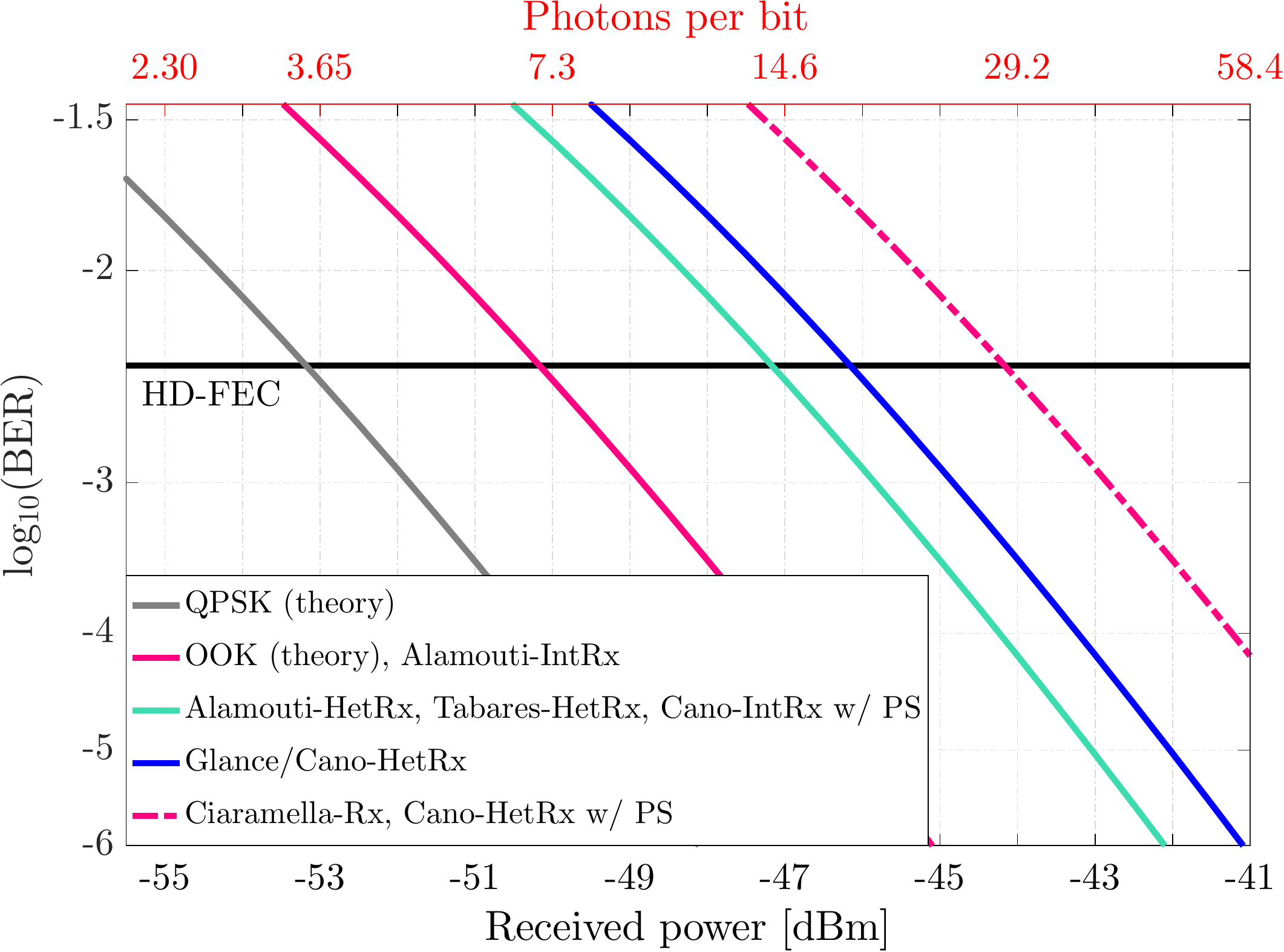}
	\caption{BER versus received power for each low complexity receiver, operating at 10.7~Gb/s. Ciaramella-Rx uses OOK signalling whereas QPSK is considered for all the other low complexity receivers.}
	\label{Fig:BERtheory_PICohRx}
\end{figure}

Alamouti-IntRx was found to require the lowest number of photons per bit (PPB), 7~PPB at the HD-FEC threshold of $4\times10^{-3}$, double the number of PPB compared to DP-QPSK in theory, as shown in Fig.~\ref{Fig:BERtheory_PICohRx}. {\color{red}Tabares-HetRx and Cano-IntRx with polarisation scrambling (PS) exhibit the same sensitivity (14~PPB) as the Alamouti-HetRx}, requiring twice the PPB compared to the Alamouti-IntRx. The energy of in-band noise in Tabares-HetRx doubles due to the detection of the noise in the orthogonal polarisation component and its addition into the signal band. Moreover, although Cano-IntRx with PS utilises intradyne detection, the PS technique comes with a 6~dB sensitivity penalty, compared to the theory, due to the sacrifice, not only of one of the polarisation modes, but also of the adjacent time slots on both polarisation modes (requiring four times the effective symbol rate compared to DP-QPSK, in theory). In contrast to the Alamouti-HetRx, Tabares-HetRx, and Cano-IntRx with PS, which all cancel the direct detection terms, the Glance/Cano-HetRx additionally suffers from the common-mode noise components, and requires 19.5~PPB at the HD-FEC threshold of $4\times10^{-3}$, as shown in Fig.~\ref{Fig:BERtheory_PICohRx}.

Finally, the Ciaramella-Rx has an intrinsic sensitivity penalty of approximately 6~dB, since the energy of in-band noise quadruples due to the addition of noise in the polarisation orthogonal to the signal, similarly to the Tabares-HetRx, and requires a large signal-LO frequency offset (close to the symbol rate), causing the image band to merge into the signal band. Note that the Ciaramella-Rx suffers from an additional 3~dB inherent sensitivity penalty compared to the Alamouti-HetRx due to its use of OOK signalling. Finally, the Cano-HetRx with PS exhibits the same performance as that of the Ciaramella-Rx,~\ie~requiring 28~PPB (two times more than the Alamouti-HetRx at the HD-FEC threshold of $4\times10^{-3}$). 

\subsection{Comparison of sensitivity limits of practical low complexity coherent receivers}\label{subsec:PerfComp}
\begin{figure}[t]
	\centering
	\includegraphics[width=1\linewidth]{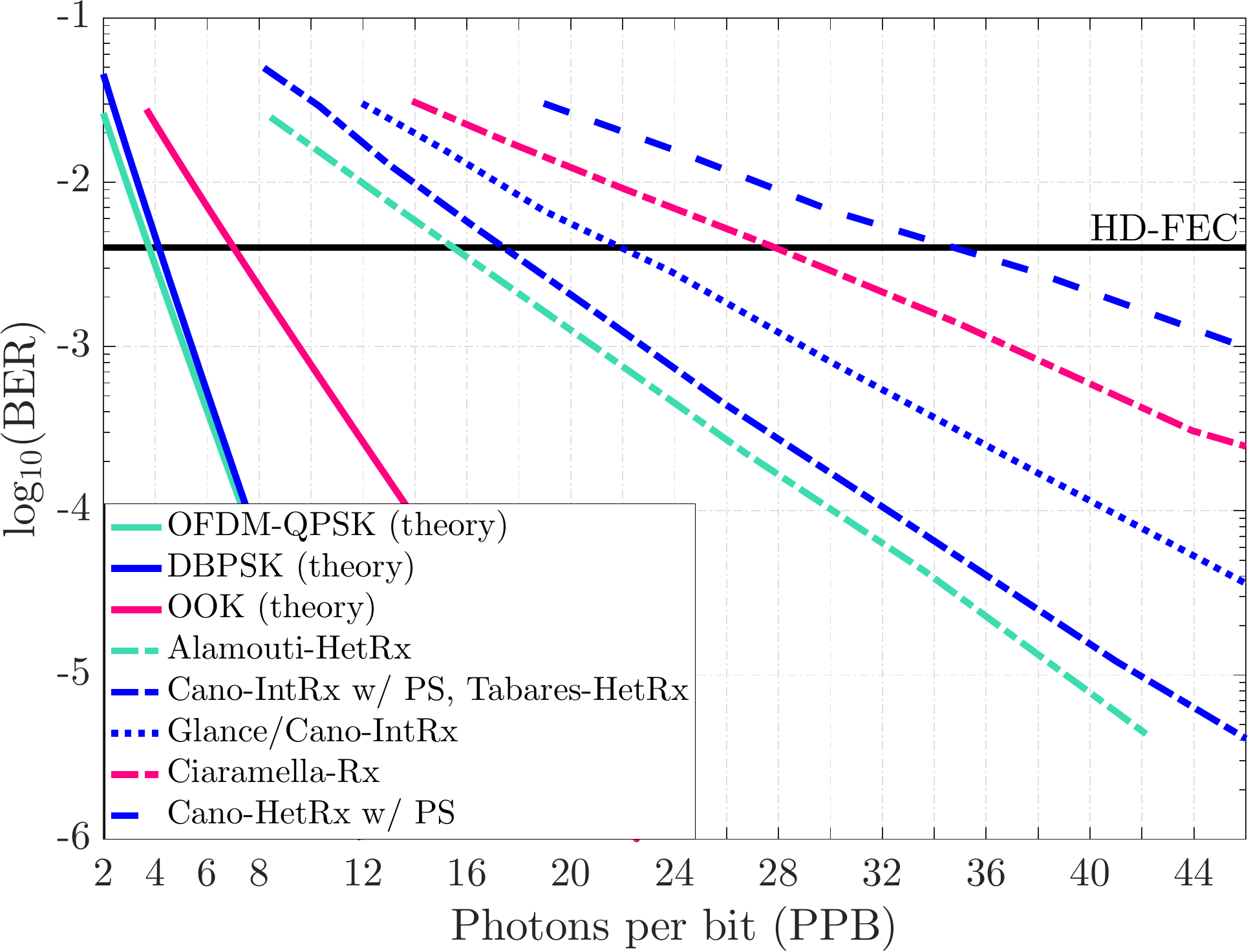}
	\caption{The sensitivity limits of the simplified coherent receivers, specified as the number of PPB, were calculated through numerical simulations using the experimentally reported parameters. The solid lines represent the shot noise limits of the modulation formats using the conventional (PPDI) coherent receiver whereas the dot-dashed and dotted lines correspond to the shot noise limits realisable using the simplified coherent receivers. PS: Polarisation scrambling}
	\label{Fig:PICohRxs_ppb}
\end{figure}

The numerical simulations were repeated considering the modulation formats used in the previously reported experimental demonstrations employing such receivers. DPSK signalling was used for Cano receivers with polarisation scrambling (PS), Glance/Cano-HetRx, and Tabares-HetRx  whereas OFDM-QPSK and OOK were considered for the Alamouti receivers and Ciaramella-Rx, respectively, and the results are presented in Fig.~\ref{Fig:PICohRxs_ppb}.

First, the theoretical receiver sensitivities for OFDM-QPSK, DBPSK and OOK signals, assuming the use of an ideal conventional PPDI coherent receiver, were calculated and are plotted in Fig.~\ref{Fig:PICohRxs_ppb} for benchmarking purposes. The sensitivity with OFDM-QPSK signalling is 3.85~PPB, which is 0.35 PPB more than with single carrier QPSK,and 0.4 and 3.15 PPB lower than with DBPSK and OOK signals, respectively. Alamouti-IntRx/HetRx requires approximately 7.8~PPB/15.5~PPB respectively, whereas the Cano-IntRx with PS and Tabares-HetRx require 17.4~PPB, and the Glance/Cano-HetRx requires 21.4~PPB. The sensitivity difference between the Alamouti-HetRx and Tabares-HetRx is solely due to the theoretical sensitivity difference between QPSK and DBPSK. Moreover, the Cano-HetRx with PS has an approximately 3 dB lower sensitivity than the Alamouti-HetRx, as shown in Fig.~\ref{Fig:PICohRxs_ppb}. The sensitivity differences between the Cano and Alamouti receivers arise from the fact that Cano receivers employ the polarisation scrambling method (sending one information symbol in two adjacent time slots) combined with the DBPSK signalling scheme whereas, the Alamouti-coding scheme (sending two information symbols in two adjacent time slots), combined with QPSK signalling, is used for the Alamouti receivers, as illustrated in Fig.~\ref{Fig:PolIndOp}. Note that, although the use of heterodyne reception comes at the price of a doubling in the required number of PPB compared to intradyne detection, it enables the simultaneous use of the ONU laser as a downstream LO laser and an upstream source laser.

The sensitivity of the Alamouti-HetRx exceeds that of the Ciaramella-Rx, requiring approximately half the PPB in ideal system implementations, as shown in Fig.~\ref{Fig:PICohRxs_ppb}. However, the main advantage of the Ciaramella-Rx low complexity coherent receiver, compared to the Alamouti-HetRx, is that the signal can be demodulated using simple, analogue processing,~\ie~ADC-less and DSP-less operation, and consequently, simplifying its real-time implementation. However, phase-diversity is not preserved due to the squaring operation, and thus, the Ciaramella-Rx exhibits low dispersion tolerance, particularly at higher bit rates, as discussed further in Section~\ref{sec:discussion}. Besides this, the realisable modulation schemes are limited to real-valued formats, such as ASK signalling. To double the spectral-efficiency compared to OOK, 4-PAM can be used, but with an associated reduction in the sensitivity (increasing the required number of PPB by approximately a factor of three), as previously indicated in Table~\ref{Tab:RequiredPPBs}. Such a limitation in the achievable spectral efficiency and the low resilience to dispersion affect the possibility of scaling the Ciaramella-Rx's operation to higher data rates. As discussed in Section~\ref{subsec:PICohRx_description}, the squaring operation in Ciaramella-Rx was removed to regain the phase diversity in the Tabares-HetRx~\cite{Tabares_OFC17}. 

\subsection{Optimizing LO power in low complexity coherent receivers}
Assuming ideal balanced detection (having an infinite CMRR) and no optical pre-amplification, the increase in LO power improves the receiver sensitivity until the system performance reaches the shot noise limit. It should be noted that, in practice, if the signal power is low (\eg~$\leq$-20~dBm), the sensitivity degrades at very high LO power values (typically $\geq$15~dBm) due to the LO relative intensity noise (RIN), giving rise to residual LO-RIN beat noise due to the finite CMRR,~\ie~imperfect balancing of the balanced photodiodes. Therefore, there exists an optimum LO power for a given RIN value, balancing the thermal noise with the residual LO-RIN  beat noise to achieve the best system performance~\cite{Zhang_OpEx12}, as discussed in Section~\ref{subsec:minCompCohRx}. However, the impact of RIN was neglected in this analysis for the sake of simplicity, and hence, the performance of the receivers converge to their shot noise limits beyond a certain LO power threshold.

\begin{figure}[t]
	\centering
	\includegraphics[width=1\linewidth]{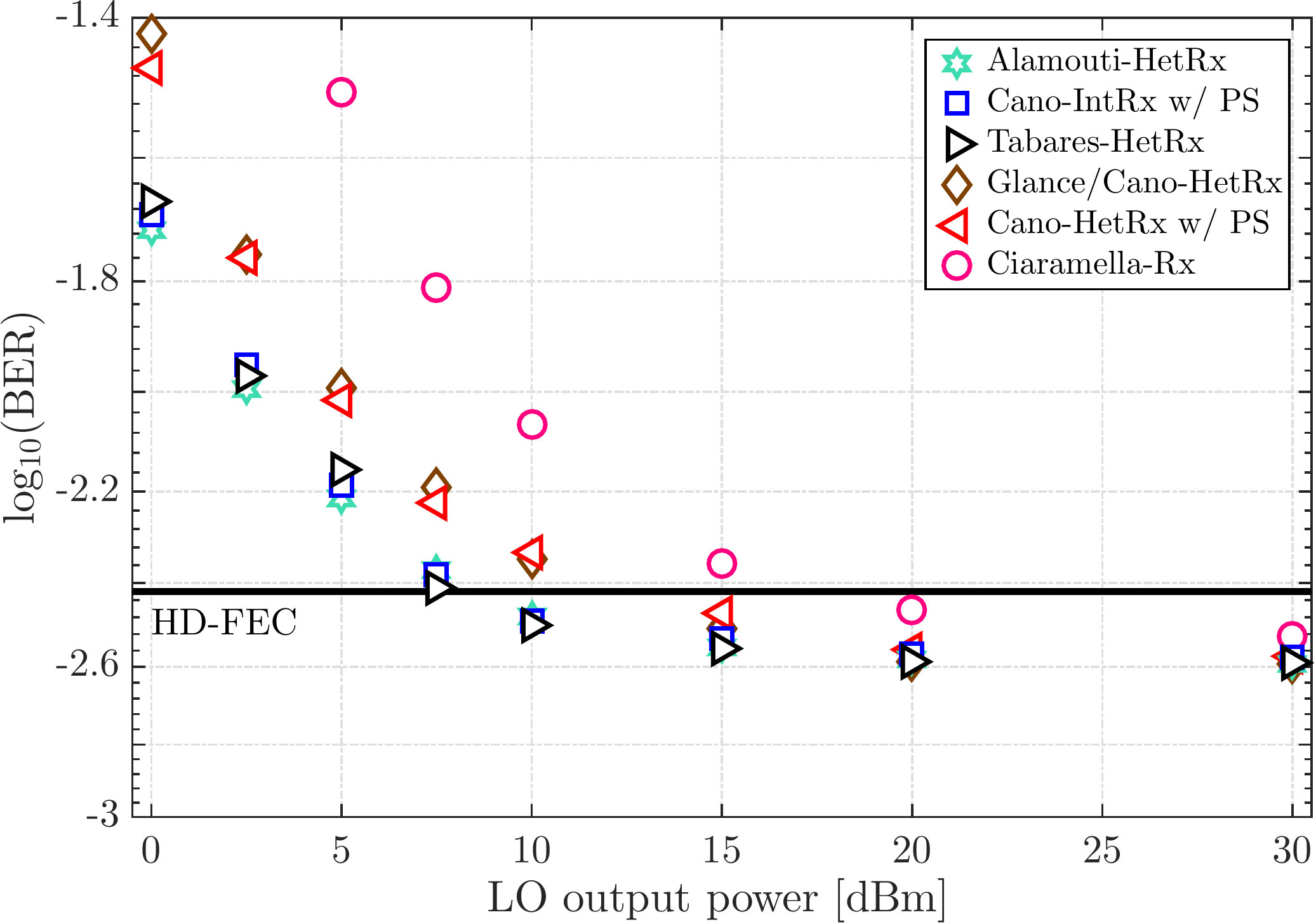}
	\caption{BER vs LO output power. For each receiver, the LO power was swept around the HD-FEC threshold BER, obtained from Fig.~\ref{Fig:PICohRxs_ppb}.}
	\label{Fig:BERvsLOpower}
\end{figure}

For each receiver, the analysis was performed at the sensitivity value (obtained from Fig.~\ref{Fig:PICohRxs_ppb}) achieving the HD-FEC threshold BER. At LO-signal power ratios greater than 20~dB, the receivers reach their shot noise limits in the absence of RIN, as expected. It is found that the Ciaramella-Rx has the greatest susceptibility to low LO power levels ($\leq$~10~dBm) due to both the use of ASK signalling and the interference from the direct detection terms, {\color{red}as shown in Fig.~\ref{Fig:BERvsLOpower}.} Glance/Cano-HetRx and Cano-HetRx with PS exhibit greater resilience compared to Ciaramella-Rx, as most of the direct detection terms appear outside the signal band. {\color{red}Finally, Alamouti-HetRx, Cano-IntRx with PS, and Tabares-HetRx exhibit the greatest ability to operate at low LO power values due to their complete removal of the direct detection terms.}

\begin{figure}[t]
	\centering
	\includegraphics[width=1\linewidth]{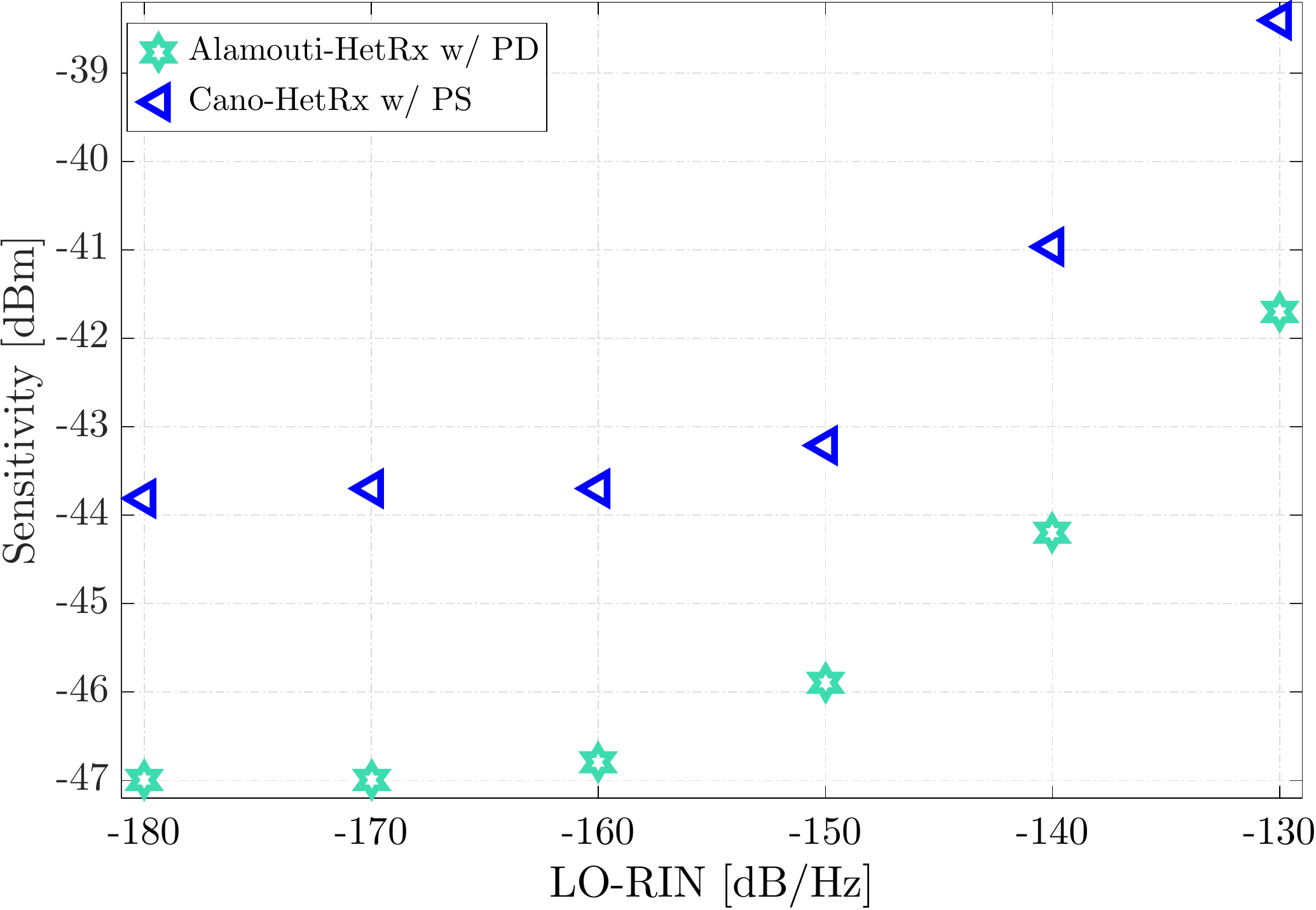}
	\caption{Receiver sensitivity versus LO-RIN and LO output power.}
	\label{Fig:OptLOpower}
\end{figure}

{\color{red}
\subsection{Comparison of Minimum Complexity (Single-ended PD-based Heterodyne) Receivers} \label{subsec:minCompCohRx}
In this section, we study the performance of minimum complexity coherent receivers, comparing the Alamouti-coding heterodyne Rx and the Cano-HetRx with PS which consist of a 3-dB coupler followed by a single-ended PD and an ADC, as depicted in Fig.~\ref{Fig:PICohRxDesigns}. An identical receiver architecture was considered in both cases, and the impact of non-negligible LO-RIN was investigated. The rest of the system parameters were kept the same, as described in Section~\ref{subsec:NumSims}. At the OLT side, the former achieves polarisation-independent operation using polarisation-time block (Alamouti) coding and a dual-polarisation modulator whereas a centralized polarisation scrambling is realised using a polarisation modulator, synchronised with the data clock running at twice the symbol rate, for the latter, as discussed in detail in Section~\ref{subsec:PICohRx_description}. Since a single-ended PD was used for detection in the presence of LO-RIN, the optimum receiver sensitivity is dictated by the trade-off between the amount of thermal noise and the beating of the LO-RIN. This results in an optimum LO power, as shown in Fig.~\ref{Fig:OptLOpower}. Note that in unamplified applications, such as PONs, since the LO power is significantly larger than the signal power, the signal-signal beating can be neglected compared to the desired LO-signal beating~\cite{Lavery_JLT13}.

For both types of receiver, it is found that the change in sensitivity with respect to LO-RIN beating is very similar at a given LO output power. In Fig.~\ref{Fig:OptLOpower}, for the RIN values of -180 and -170~dB/Hz, the minimum LO output power, achieving the pre-FEC BER of $4\times10^{-3}$, is found to be 10~dBm. There is no notable sensitivity penalties observed up to -160~dB/Hz RIN. However, penalties of 1, 2.9 and 5.4~dB were observed for -150,-140 and -130~dB/Hz RIN at the optimum LO output power of 8,5 and 0~dBm respectively, as shown in Fig.~\ref{Fig:OptLOpower}. It should be noted that the impact of signal-signal beating is insignificant since both receivers use heterodyne detection,~\ie~most of the direct detection terms appear outside of the signal band, at high LO-to-signal power ratio. Alamouti-HetRx with a single-ended PD outperforms Cano-HetRx with PS by 3~dB, solely due to the difference in realisation of polarisation-independent operation, as discussed in Section~\ref{subsec:PerfComp}, at the expense of using a dual-polarisation modulator (instead of a synchronous polarisation scrambler) in the OLT.}
% * <r.killey@ucl.ac.uk> 2018-02-09T12:52:09.839Z:
% 
% In Fig. 10, plot the optimum LO-signal power on the y2 axis, as a separate curve. Use arrows from the curves to point to the corresponding y and y2 axes. Plot the values as LO-signal power ratio in dB (not LO power in dBm)
% 
% ^.

\begin{table*}[t] 
	\small
	\centering
	\captionsetup{name=Table}
	\caption{Performance and optical complexity comparison of the simplified coherent receivers achieving polarisation independent detection. The theoretical shot noise limits for B2B required PPB (obtained via numerical simulations) and reported experimental demonstrations listing the achieved sensitivities and power budgets at a given transmission distance and bit rate at the HD-FEC threshold (assumed to be $4\times10^{-3}$) are presented.} 
	\label{Tab:PICohRxComparison}
	\renewcommand{\arraystretch}{1.1}% Tighter
	\newcolumntype{C}[1]{>{\centering\arraybackslash}m{#1}}
	\begin{center}
		\vspace{-1pc}
		\resizebox{\textwidth}{!}
        {
			\begin{tabular}{ C{3.15cm} C{1.65cm} C{2.75cm} C{3cm} C{2.3cm} C{3cm} }
				
             \toprule
				Simplified Coh. Rx &  \makecell{Modulation \\ (SE [bit/s/Hz])} & \makecell{B2B req. PPB in \\ sim. (see Fig.~\ref{Fig:PICohRxs_ppb})} & Exp. sensitivity/Bit rate [{\color{red}PPB}]/[Gbps]  & Distance [km] &  ONU Rx \\ [1ex]
				\midrule \midrule 
				
				%\parbox[c][0cm][c]{3.2cm}{}
				Ciaramella-Rx\cite{Artiglia_JLT15} &  OOK (1) & 28 & 246.8/10 & 105 \hspace{3cm} (bidirectional) & \makecell{PBS + 3$\times$3 coupler \\ + 3 PDs} \\ [0.5ex] 
				
				%\parbox[c][0cm][c]{3.2cm}{}
				Tabares-Rx\cite{Tabares_OFC17} &  DBPSK (1) & 17.5 & 78.5/1.25 & 50 (not \hspace{2cm} bidirectional) & \makecell{PBS + 3$\times$3 coupler \\ + 3 PDs}\\ [0.5ex] 
				
				Cano-IntRx w/ PS\cite{Cano_ECOC15} &  DBPSK (1) & 17.5 & 78.5/1.25 & 50 (not \hspace{2cm} bidirectional) & \makecell{3$\times$3 coupler \\ + 3 PDs} \\ [0.5ex] 
					
				Cano-HetRx w/ PS\cite{Cano_ECOC14} &  DBPSK (1) & 34.7 & 197.4/1.25 & 50 (not \hspace{2cm} bidirectional) &  3-dB coupler + 1 PD \\ [0.5ex] 
				
				Glance/Cano-IntRx\cite{Cano_ECOC15-Glance}  &  DBPSK (1) & 21.9 & 123.6/10 & 25 (not \hspace{2cm} bidirectional) & 3-dB coupler +  1 PBS + 2 PDs + 2 ADCs \\ [0.5ex] 
				
 				\rowcolor[gray]{0.9}
				%\parbox[c][0cm][c]{3.5cm}{} %{\hspace{-.3cm}
				Alamouti-HetRx\cite{Erkilinc_NatureComm17}  &  \makecell{AC-OFDM \\ QPSK (2)} & 15.5 & 58/10.7 & 108  \hspace{3cm} (bidirectional)  & 3-dB coupler + 1 BPD + 1 ADC \\ [0.5ex] 
				
				\rowcolor[gray]{0.9}
				%\parbox[c][0cm][c]{3.5cm}{} %\hspace{-.3cm}
				Alamouti-HetRx\cite{Erkilinc_NatureComm17}  &  \makecell{AC-OFDM \\ 16QAM (4)} & 36.5 & 230/21.4 & 38 \hspace{3cm} (bidirectional)  & 3-dB coupler + 1 BPD + 1 ADC \\ [0.5ex] 
				
				\bottomrule
				\multicolumn{6}{c}{\parbox[t][0.85cm][c]{18cm}{AC:Alamouti-coded. SE:Spectral efficiency. B2B:Back-to-back. Req. PPB:Required photons per bit. PD:Photodiode. BPD:Balanced photodiode. PBS:Polarisation beam splitter. ADC:Analogue-to-digital converter.}}
       
		\end{tabular}
        }
	\end{center}
	\vspace{-1pc}
\end{table*}

\section{Experimentally Achieved Receiver Sensitivities Using Low Complexity Coherent Receivers} \label{sec:discussion}
The sensitivity limits of low complexity coherent receivers obtained in ideal system simulations and in experiments measuring back-to-back receiver sensitivities (in PPB), power budgets (in dB) and transmission reach are summarised in Table~\ref{Tab:PICohRxComparison}. The table includes lists of the required receiver components. The systems operating at 1.25~Gb/s using the Cano-IntRx and Cano-HetRx with polarisation scrambling (PS) were demonstrated, with reported receiver sensitivities of -49~dBm (78.5~PPB) and -45~dBm (197.4~PPB), respectively, at a pre-FEC BER of $10^{-3}$ over a transmission distance of 50~km ~\cite{Cano_ECOC15,Cano_ECOC14}. Furthermore, the bidirectional real-time implementation of the Glance/Cano-IntRx (using only simple analogue processing rather than DSP in the receiver) operating at 1.25~Gb/s with the DBPSK signal format was demonstrated in a field trial over 10~km SSMF, exhibiting a sensitivity of -37.5~dBm (1110~PPB)~\cite{Presi_JLT17}. There is no reported experimental demonstration operating at 10~Gb/s using the Cano-IntRx/HetRx with PS whereas the Glance/Cano-IntRx was used in a 10~Gb/s system, achieving a sensitivity of -38~dBm (123.6~PPB), over a transmission distance of 25~km~\cite{Cano_ECOC15-Glance}. 

Moreover, the Tabares-HetRx was employed in an ultra-dense WDM PON system operating at 1.25~Gb/s using DBPSK signalling, exhibiting a sensitivity of -49~dBm (78.5~PPB) at the HD-FEC threshold of $10^{-3}$~\cite{Tabares_OFC17}. The real-time implementation of Ciaramella-Rx, using simple analogue processing without the need for DSP or ADC, was demonstrated for a 1.25~Gb/s~OOK PON system, achieving a sensitivity of −51~dBm (49.6~PPB) at a pre-HD-FEC BER of $2\times10^{-3}$~\cite{Artiglia_JLT16}. The same receiver in bidirectional operation was also tested in a field trial using 35~km SSMF, achieving a sensitivity of -48~dBm (99~PPB) at a HD-FEC threshold of $1.8\times10^{-3}$~\cite{Presi_JLT17}. The demonstration of a 10~Gb/s coherent-enabled WDM-PON system using the Ciaramella-Rx was reported in~\cite{Artiglia_JLT15} in which a receiver sensitivity of -38~dBm (123.6~PPB) was achieved at the HD-FEC threshold of $2\times10^{-3}$ in back-to-back operation. However, its sensitivity was affected significantly after transmission, due to its low resilience to chromatic dispersion caused by dispersion-induced power fading upon photodetection. In particular, a 4~dB sensitivity penalty, leading to a receiver sensitivity of -34~dBm (310.5~PPB), at the HD-FEC threshold of $2\times10^{-3}$  was observed following transmission over a distance of 52~km. To overcome this limitation due to dispersion, a chirp-managed laser was used as a source laser in the downstream transmitter, and more importantly, three ADCs followed by three 4$^{\text{th}}$-order Gaussian-shaped band-pass electrical filters were added in the receiver side~\cite{Rannelo_OpEx17}. The penalty due to the chromatic dispersion was reduced from 4~dB to 1.5~dB. By applying such modifications, a coherent-enabled 10~Gb/s PON system solution offering a sensitivity of -35~dBm (246.8~PPB) over a transmission distance of 105~km was achieved. However, a second laser or a photonic mixer to shift the wavelength of a laser in order to be used as an upstream signal source laser is required for bidirectional operation.

Finally, a receiver sensitivity of -41~dBm (58~PPB) using the Alamouti-HetRx in back-to-back operation was recently achieved, with bidirectional transmission demonstrated over installed fibre links of up to 108~km, operating at a symmetrical bit rate of 10.7~Gb/s. The ONU laser employed in these experimental demonstrations was simultaneously used as both the LO and source lasers to detect the downstream and to generate the upstream signals, respectively. The key benefits of the Alamouti-HetRx are that it has a complexity comparable to currently employed direct detection receivers whilst requiring just 58~PPB (-41~dBm receiver sensitivity), as demonstrated in~\cite{Erkilinc_NatureComm17} whereas currently employed direct-detection receivers without optical pre-amplification (using OOK formats) require in the range of thousands of PPB. The achievable bit rates using the Alamouti-HetRx are scalable as the phase-diversity is preserved, in contrast to the Ciaramella-Rx, leading to higher achievable spectral-efficiencies, as presented in Table~\ref{Tab:PICohRxComparison}. Moreover, a sensitivity penalty of just 0.5~dB was incurred due to the use of a 5\% cyclic prefix overhead to achieve dispersion tolerance in a 108~km standard SMF link.

\section{Conclusions}
A variety of low complexity (simplified) coherent receiver designs, proposed to date for use in the optical network units of access networks, has been comprehensively investigated and compared in terms of their complexity and achievable sensitivities. To the best of our knowledge, this paper is the first to report a detailed side-by-side comparison of such receivers.

Our analysis indicates that the Ciaramella-Rx and Glance/Cano-HetRx are favourable for ADC-less and DSP-less operation in an ONU for systems operating in the range of 1 to 5~Gb/s. However, both of these receivers require a polarisation beam splitter, known to be challenging to monolithically integrate. Further, only real-valued signalling such as ASK is realisable using the Ciaramella-Rx, which limits the achievable spectral efficiency, and requires a second laser to generate an upstream signal in the ONU, whereas the Glance/Cano-HetRx requires a DAC to achieve higher data rates ($\geq$5~Gb/s). 

{\color{red}Assuming the use of single carrier QPSK signalling, the recently proposed Alamouti-HetRx, Tabares-HetRx and Cano-IntRx with PS offer the highest sensitivities, 15.5~PPB, compared to other low complexity receivers. However, Cano-IntRx with PS and Tabares-HetRx require significantly higher optical complexity compared to Alamouti-HetRx. Besides this, Cano-IntRx with PS requires an additional laser to generate the upstream signal as opposed to Alamouti-HetRx and Tabares-HetRx. The complexity difference between Tabares-HetRx and Alamouti-HetRx comes from the fact that polarisation-independent operation is enabled by the coding scheme applied in the transmitter whereas with the Tabares-HetRx, this is achieved using all three input ports of a symmetric 3$\times$3 coupler combined with a PBS. Moreover, the Alamouti-HetRx exhibits 5 and 14~PPB less than the Glance/Cano-HetRx (which requires an additional PBS before a 3-dB coupler) and Cano-HetRx with PS, respectively. Compared to the currently employed direct detection receivers, Alamouti-HetRx and Cano-HetRx with PS offer comparable optoelectronic complexity.}

Moreover, Alamouti-HetRx has been used to demonstrate the highest experimental sensitivity (58~PPB), achieving the highest bit rate per $\lambda$, in bidirectional transmission to date. The key importance of this demonstration is the realisation of bandwidth-efficient modulation formats and electronic chromatic dispersion compensation, as well as the simultaneous use of the ONU laser as downstream LO and and upstream signal source, enabled by heterodyne detection. {\color{red} It should be noted that Alamouti-HetRx comes at the expense of higher DSP complexity. Besides the DSP complexity, the required optical complexity in the transmitter compared to Tabares-HetRx, Glance/Cano-HetRx and Cano-HetRx with PS is higher for the Alamouti-HetRx.} Nonetheless, such complexity may become insignificant considering the potential reduction in receiver bandwidth requirements due to both frequency selectivity and advanced modulation, as well as the increase in the number of subscribers supported by the network. It can be concluded that, depending on the desired system capacity and reach, the proposed simplified coherent ONU transceivers, as opposed to direct detection ONU transceivers, have the potential to offer promising low cost solutions for future coherent-enabled WDM-PON (\eg~optical access and mobile backhaul network) applications.

\section*{Acknowledgements}
This work was supported by the EPSRC Programme Grant UNLOC EP/J017582/1, EP/J008842/1 and Huawei Technologies. Dr. Domani\c{c} Lavery is supported by the Royal Academy of Engineering under the Research Fellowships scheme.


\begin{thebibliography}{12}
	\bibitem{Cisco17}
	Cisco, ``Cisco visual networking: Forecast and methodology 2016-2021,'' 2017. 
    % (date of access: 16/11/2014), URL http://www.cisco.com/c/en/us/solutions/collateral/service-provider/visual-networking-index-vni/complete-white-paper-c11-481360.pdf.
	
    \bibitem{EC}
    European Commission, ``The EU explained: Digital agenda for Europe,'' 2017.
    
    \bibitem{Huawei13}
	Huawei, ``Next-generation PON evolution,'' 2013 (date of access: 16/10/2017), URL 
	www.huawei.com/ilink/en/download/HW\_077443.
        
    \bibitem{Nesset_JOCN17} 
    D.~Nesset, ``PON Roadmap [Invited],'' \textit{J. Opt. Commun. Netw.,} vol.~9, no.~1, pp.A71--A76, 2017.
    
    \bibitem{ITU_NG-PON2}
	ITU, ``ITU-T recommendation G.989 40-Gigabit-capable passive optical networks (NG-PON2): General requirements,'' 2013 (date of access: 16/10/2017), URL https://www.itu.int/rec/T-REC-G.989.1/e.
    
	\bibitem{Kani10}
	J.-I.~Kani,  ``Enabling technologies for future scalable and flexible WDM-PON and WDM/TDM-PON systems,'' \textit{IEEE J. Selected Top. in Quant. Electron.}, vol.~16, no.~5, pp.~1290--1297, 2010.  
		
	\bibitem{Pachnicke16}
	S.~Pachnicke, J.~Zhu, M.~Lawin, M.H.~Eiselt, S.~Mayne, B.~Quemeneur, D.~Sayles, H.~Schwuchow, A.~Wonfor, P.~Marx, M.~Fellhofer, P.~Neuber, M.~Dietrich, M.J.~Wale, R.V.~Penty, I.H.~White, J.-P.~Elbers, ``Tunable WDM-PON system with centralized wavelength control,'' \textit{J. Lightw. Technol.}, vol.~34, no.~~2, pp.~812--818, 2016.
	
    \bibitem{ITU_XGPON}
	ITU, ``ITU-T Recommendation G.987 10-Gigabit-capable passive optical network (XG-PON) systems,'' 2012 (date of access: 16/10/2017), URL https://www.itu.int/rec/T-REC-G.987.3/en.

	\bibitem{IEEE_EPON}
	IEEE, ``IEEE P802.3av10 Gb/s Ethernet passive optical network (EPON)'', 2009 (date of access: 16/10/2017), URL http://www.ieee802.org/3/av/.
    
    %\bibitem{LookPachnicke16}
    %Next generation broadband in Europe: The need for speed. \textit{HeavyReading Report}, vol.~3, 2005.
    
    \bibitem{Bhar14} 
    C.~Bhar, G.~Das, A.~Dixit, B.~Lannoo, D.~Colle, M.~Pickavet, and P.~Demeester, ``A novel hybrid WDM/TDM PON architecture using cascaded AWGs and tunable components,'' \textit{J. Lightw. Technol.,} vol.~32, no.~9, pp.~1708--1716, 2014.

	\bibitem{Grobe_JLT14} 
    K.~Grobe, M.H.~Eiselt, S.~Pachnicke and J.-P.~Elbers, ``Access Networks Based on Tunable Lasers,'' \textit{J. Lightw. Technol.,} vol.~32, no.~16, pp.~2815--2815, 2014.
    
	\bibitem{Lavery_OpEx10}
	D.~Lavery, M.~Ionescu, S.~Makovejs, E.~Torrengo, and S.J.~Savory,    ``A long-reach ultra-dense 10 Gbit/s WDM-PON using a digital  coherent  receiver,'' \textit{Opt. Exp.}, vol.~18, no.~25, pp.~25855--25860, 2010.
	
	\bibitem{Ferreira_PTL17}
	R.~M.~Ferreira, A.~Shahpari, J.~D.~Reis, and A.~L.~Teixeira, ``Coherent UDWDM-PON with dual-polarization transceivers in real-time," \textit{IEEE Photon. Technol. Lett.,} vol.~29, no.~11, pp.~909--912, 2017.  
	
	%   Field-trial of a real-time bidirectional UDWDM-PON coexisting with GPON, RF video overlay and NG-PON2 systems. \textit{Paper presented at European Conference on Optical Communication} (ECOC), Valencia (Spain), PDP.4.5 (2015). DOI:10.1109/ECOC.2015.7341693.
	
	\bibitem{Rohde_JLT14}
	H.~Rohde, E.~Gottwald, A.~Teixeira, J.D.~Reis, A.~Shahpari, K.~Pulverer and J.S.~Wey, ``Coherent ultra dense WDM technology for next generation optical metro and access networks,'' \textit{J. Lightw. Technol.}, vol.~32, no.~10, pp.~2041--2052, 2014. 

	\bibitem{Shahpari17}
A.~Shahpari, R.~M.~Ferreira, R.~S.~Luis, Z.~Vujicic, F.~P.~Guiomar, J.~D.~Reis, A.~L.~Teixeira, ``Coherent access: A review,'' \textit{J. of Lightw. Technol.,} vol.~35, no.~4, pp.~1050--1058, 2017,

	\bibitem{Shea_JLT07}
	 D.~P.~Shea and J.~E.~Mitchell, ``A 10-Gb/s 1024-way-split 100-km long-reach optical access network,'' \textit{J. Light. Technol.}, vol.~25, no.~3, pp.~685--693, 2007.
    
	\bibitem{Jensen_JLT14}
	J.~B~Jensen, ``VCSEL Based Coherent PONs," \textit{J. Lightw. Technol.}, vol.~32, no.~8, pp.~1423-1433, 2014. 
	
    \bibitem{Proakis08}
    J.~G.~Proakis and M.~Salehi,  ``Digital communications,'' 5th ed., McGraw-Hill, 2008.
    
    \bibitem{Erkilinc_ECOC15}
	M.~S.~Erk{\i}l{\i}n\c{c}, D.~Lavery, R.~Maher, M.~Paskov, B.C.~Thomsen, R.I.~Killey, P.~Bayvel, and S.J.~Savory, ``Polarization-insensitive single balanced photodiode coherent receiver for passive optical networks,'' in \textit{Proc. European Conference on Optical Communication (ECOC)}, paper Th.1.3.3, 2015. 
    
    \bibitem{Glance87}
    B.~Glance, ``Polarization independent coherent optical receiver,'' \textit{J. Lightw. Technol.}, vol.~5, no.~2, pp.~274--276, 1987.
    
	\bibitem{Cano_ECOC15}
	I.~N.~Cano, A.~Lerin, V.~Polo and J.~Prat, ``Flexible D(Q)PSK 1.25-5 Gb/s UDWDM-PON with directly modulated DFBs and centralized polarization scrambling,'' in \textit{Proc. European Conference on Optical Communication (ECOC)}, paper Th.1.3.7, 2015. 
	
	\bibitem{Cano_ECOC14}
	I.~N.~Cano, A.~Lerin, V.~Polo and J.~Prat, ``Polarization independent single-PD coherent ONU receiver with centralized scrambling in udWDM-PONs,'' in \textit{Proc. European Conference on Optical Communication (ECOC)}, paper P.7.12, 2014.
    
    \bibitem{Ciaramella_PTL14}
	E.~Ciaramella, ``Polarization-independent receivers for low-cost coherent OOK systems,'' \textit{IEEE Photon. Technol. Lett.}, vol.26, no.~6, pp.~548--551, 2014.
	
	\bibitem{Tabares_OFC17}
	J.~Tabares, V.~Polo and J.~Prat, ``Polarization-independent heterodyne DPSK receiver based on 3x3 coupler for cost-effective udWDM-PON,'' in	\textit{Proc. Optical Fiber Communication Conference (OFC)}, paper Th1K.3, 2017. 
    
    \bibitem{Agrawal05}
	G.~P.~Agrawal, ``Lightwave Technology: Telecommunication Systems, John Wiley \& Sons, 2005.

	\bibitem{Liu08}
    I.~P.~Kaminow, T.~Li and A.~E.~Willner, ``Optical fiber telecommunications VB: Systems and networks,'' Academic Press, 2008.
    
	\bibitem{Liu11}
	 X.~Liu, S.~Chandrasekhar, T.~H.~Wood, R.~W.~Tkach, P.~J.~Winzer, E.~C.~Burrows, and
 A.~R.~Chraplyvy, ``M-ary pulse-position modulation and frequency-shift keying with additional polarization/phase modulation for high-sensitivity optical transmission,'' \textit{Opt. Express}, vol.~19, no.~26, pp.~B868--881, 2011.
	
	\bibitem{Ludwig15}
	A.~Ludwig, M.~-L.~Schulz, P.~Schindler, S.~Wolf, C.~Koos, W.~Freude and J.~Leuthold, ``Stacked modulation formats enabling highest-sensitivity optical free-space links,'' \textit{Opt. Express}, vol.~23, no.~17, pp.~21942--21957, 2015.	
		
	\bibitem{Agrell09}
	E.~Agrell and M.~Karlsson, ``Power-efficient modulation formats in coherent transmission systems,'' \textit{J. Lightw. Technol.}, vol.~27, no.~22, pp.~5115--5126, 2009.

	\bibitem{Lavery_OpEx11}
	D.~Lavery,  C.~Behrens, and S.~J.~Savory, ``A comparison of modulation formats for passive optical networks,'' \textit{Opt. Express} vol.~19, no.~26, pp.~B836--B841, 2011.
    
    \bibitem{Shieh08}
    W.~Shieh, ``Coherent optical OFDM: has its time come?,'' \textit{J. Opt. Netw.}, vol.~7, no.~3, pp.~234--255, 2008.
    
    \bibitem{Erkilinc_JLT16}
    M.S.~Erk{\i}l{\i}n\c{c}, D.~Lavery, B.C.~Thomsen, R.I.~Killey, P.~Bayvel, and S.J.~Savory, ``Polarization-insensitive single-balanced photodiode coherent receiver for long-reach WDM-PONs,'' \textit{J. Lightw. Technol.} vol.~34, no.~8, pp.~2034--2041, 2016.
    
    \bibitem{Faruk_OpEx16}
     Md.S.~Faruk, H.~Louchet, M.S.~~Erk{\i}l{\i}n\c{c}, and S.J.~Savory, ``DSP algorithms for recovering single-carrier Alamouti coded signals for PON applications,'' \textit{Opt. Express}, vol.~24, no.~21, pp.~24083--24091, 2017.
    
    \bibitem{Erkilinc_ECOC16}
	M.~S.~Erk{\i}l{\i}n\c{c}, D.~Lavery, B.C.~Thomsen, R.I.~Killey, P.~Bayvel, and S.J.~Savory, ``Bidirectional symmetric 8$\times$10.7~Gb/s WDM-PON over 108~km installed fiber using low complexity polarization-insensitive coherent ONUs,'' in \textit{Proc. European Conference on Optical Communication} (ECOC), paper M.1.E.2, 2016. 
    
    \bibitem{Erkilinc_NatureComm17}
	M.~S.~Erk{\i}l{\i}n\c{c}, D.~Lavery, B.C.~Thomsen, R.I.~Killey, S.J.~Savory, and P.~Bayvel, ``Bidirectional wavelength division multiplexed transmission over installed fibre using a simplified optical coherent access transceiver,'' \textit{Nature Communications}, vol.~8, no.~1043, 2017.
        
%    \bibitem{Cano_ECOC15-Glance}
%     Cano, I. \textit{et. al.}, Directly modulated DFB with phase diversity in time polarization independent intradyne receiver for UDWDM-PON. \textit{Paper presented at European Conference on Optical Communication} (ECOC), Valencia (Spain), Th.1.3.4 (2015). DOI: 10.1109/ECOC.2015.7341845.
    
    \bibitem{Cano_ECOC15-Glance}
    I.~N.~Cano, J.C.~Velasquez, V.~Polo and J.~Prat, ``10 Gbit/s phase time diversity directly modulated DFB with single-PD intradyne receiver for coherent WDM-PON,'' in\textit{Proc. European Conference on Optical Communication (ECOC)} , paper W.4.P1.SC7.74, 2016.
    
    \bibitem{Zhou94}
	J.~Zhou and N.~Caponio, ``Operative characteristics and application aspects of synchronous intra-bit polarization spreading for polarization independent heterodyne detection'' \textit{IEEE Photon. Technol. Lett.}, vol.~6, no.~3, pp.295--298, 1994.
    
    \bibitem{Xie_ECOC11}
    C.~Xie,  P.~Winzer, G.~Raybon, A.~Gnauck, B.~Zhu, T.~Geisler, and B.~Edvold, ``Colorless coherent receiver using 3x3 coupler hybrids and single-ended detection,'' in {Proc. of European Conference on Optical Communication (ECOC)}, paper Th.13.B.2, 2011. 
        
    \bibitem{Agrawal10}
    G.P.~Agrawal, ``Fiber-Optic communication systems,'' 4th ed., John Wiley \& Sons, 2010.
       
    \bibitem{Kikuchi_JLT08}
	K.~Kikuchi, and S.~Tsukamoto, ``Evaluation of sensitivity of the digital coherent receiver,'' \textit{J. Lightw. Technol.}, vol.~26, no.~13, pp.~1817--1822, 2008.

    \bibitem{Zhang_OpEx12}
    B.~Zhang,  C.~Malouin, and T. J.~Schmidt, ``Design of coherent receiver optical front end for unamplified applications,'' \textit{Opt. Exp.}, vol.~20, no.~3, pp.~3225–-3234, 2012.

	\bibitem{Lavery_JLT13}
	D. Lavery, R.~Maher, D.S.~Millar, B.C.~Thomsen, P.~Bayvel, and S.J.~Savory, ``Digital coherent receivers for long-reach optical access networks,'' \textit{J. Lightw. Technol.}, vol.~31, no.~4, pp.~609--620, 2013.
    
	\bibitem{Presi_JLT17}
    M.~Presi, M.~Artiglia, F.~Bottoni, M.~Rannello, I.N.~Cano, J.~Tabares, J.-C.~Velásquez, S.~Ghasemi, V.~Polo, G.Y.~Chu, J.~Prat, G.~Azcarate, R.~Pous, C.~Vilá, H.~Debregeas, Gemma Vall-llosera, A.~Rafel, E.~Ciaramella, ``Field-trial of a high-budget, filterless, $\lambda$-to-the-user, UDWDM-PON enabled by an innovative class of low-cost coherent transceivers,'' \textit{J. Lightw. Technol.}, vol.~35, no.~23, pp.~5250--5259, 2017.

    \bibitem{Artiglia_JLT16}
	M.~Artiglia, M.~Presi, F.~ Bottoni, M.~Rannello and E.~Ciaramella, ``Polarization-independent coherent real-time analog receiver for PON access systems,'' \textit{J. Lightw. Technol.}, vol.~34, no.~8, pp.~2027--2033, 2016.

	\bibitem{Artiglia_JLT15}
	M.~Artiglia, R.~Corsini, M.~Presi, F.~Bottoni, G.~Cossu and E.~Ciaramella, ``Coherent systems for low-cost 10 Gb/s optical access networks,'' \textit{J. Lightw. Technol.}, vol.~ 33, no.~15, pp.~3338--3344, 2015. 
    
    \bibitem{Rannelo_OpEx17}
     M.~Rannello, M.~Artiglia, M.~Presi, and E.~Ciaramella, ``10 Gb/s long-reach PON system based on directly modulated transmitters and simple polarization independent coherent receiver,'' \textit{Optics Express}, vol.~25, no.~15, pp.~17841--17846, 2017.
       
% 	\bibitem{Dora_OFC11}
% 	Veen, D.v. \textit{et. al.} Demonstration of a Symmetrical 10/10 Gbit/s XG-PON2 System. \textit{Presented at Optical fibre Communication} (OFC) conference, Los Angeles (USA), NTuD.2 (2012). DOI:10.1364/NFOEC.2011.NTuD2.
    
%	\bibitem{Nesset15}
%	Nesset, D. NG-PON2 technology and standards. \textit{J. Lightw. Technol.} \textbf{33}, 1136--1143 (2015).
%				
%	\bibitem{Doerr12}
%	Doerr, C.R. \textit{et. al.} Monolithic polarization and phase diversity coherent receiver in silicon. \textit{J. Lightw. Technol.} \textbf{28}, 520--525 (2012).
%	
%	\bibitem{Verbist16}
%	Verbist, J. \textit{et. al.} A 40-GBd QPSK/16-QAM integrated silicon coherent receiver.\textit{IEEE Photon. Technol. Lett.} \textbf{28}, 2070--2073 (2016).

\end{thebibliography}
\end{document}